\documentclass[a4paper,11p]{article}
\pdfoutput=1 % if your are submitting a pdflatex 
\usepackage{jcappub}
\usepackage{amsmath,amsfonts,amssymb,mathtools}
\usepackage{color}
\usepackage{tikz}
\usepackage{comment}
\usepackage{multirow}

\renewcommand{\vec}{\mathbf}

\newcommand{\be}{\begin{equation}}
\newcommand{\ee}{\end{equation}}
\newcommand{\bea}{\begin{eqnarray}}
\newcommand{\eea}{\end{eqnarray}}
\newcommand{\bdm}{\begin{displaymath}}
\newcommand{\edm}{\end{displaymath}}
\newcommand{\nn}{\nonumber}

\newcommand{\xv}{\mathbf{x}}

\newcommand{\kv}{\mathbf{k}}
\newcommand{\nv}{\mathbf{n}}

\newcommand{\qv}{\mathbf{q}}
\newcommand{\pv}{\mathbf{p}}

\newcommand{\Mpc}{\, h^{-1} \, {\rm Mpc}}
\newcommand{\cMpc}{\, h^{-3} \, {\rm Mpc}^3}

\newcommand{\cGpc}{\, h^{-3} \, {\rm Gpc}^3}
\newcommand{\kMpc}{\, h \, {\rm Mpc}^{-1}}

\newcommand{\pin}{{\sc{Pinocchio}}}

\definecolor{ForestGreen}{rgb}{0.13, 0.55, 0.13}
\definecolor{airforceblue}{rgb}{0.36, 0.54, 0.66}
\definecolor{orange}{rgb}{1.0, 0.5, 0.0}
\definecolor{amethyst}{rgb}{0.6, 0.4, 0.8}
\definecolor{awesome}{rgb}{1.0, 0.13, 0.32}
\definecolor{chromeyellow}{rgb}{1.0, 0.65, 0.0}

\title{Bispectrum-window convolution via Hankel transform}

\author[a,b,c,d,e,1]{Kevin Pardede,\note{Corresponding author.}}
\author[e,f,c]{Federico Rizzo,}
\author[c,a,e,d]{Matteo Biagetti,}
\author[g]{Emanuele Castorina,}
\author[e,d,c]{Emiliano Sefusatti,}
\author[e,f,c]{and Pierluigi Monaco}

\affiliation[a]{SISSA - International School for Advanced Studies, Via Bonomea 265, 34136 Trieste,  Italy}
\affiliation[b]{ICTP, International Centre for Theoretical Physics, Strada Costiera 11, 34151, Trieste, Italy}
\affiliation[c]{Institute for Fundamental Physics of the Universe, Via Beirut 2, 34151 Trieste, Italy}
\affiliation[d]{Istituto Nazionale di Fisica Nucleare, Sezione di Trieste,  via  Valerio  2,  34127 Trieste,  Italy}
\affiliation[e]{Istituto Nazionale di Astrofisica, Osservatorio Astronomico di Trieste, via Tiepolo 11, 34143 Trieste, Italy}
\affiliation[f]{Dipartimento di Fisica, Sezione di Astronomia, Universit\`a di Trieste, via Tiepolo 11, 34143 Trieste, Italy}
\affiliation[g]{Dipartimento di Fisica `Aldo Pontremoli', Universit\`a degli Studi di Milano, Via Celoria 16, 20133 Milan, Italy}
\emailAdd{kpardede@sissa.it}

\abstract{We present a method to perform the exact convolution of the model prediction for bispectrum multipoles in redshift space with the survey window function. We extend a widely applied method for the power spectrum convolution to the bispectrum, taking advantage of a 2D-FFTlog algorithm. As a preliminary test of its accuracy, we consider the toy model of a spherical window function in real space. This setup provides an analytical evaluation of the 3-point function of the window,  and therefore it allows to isolate and quantify possible systematic errors of the method. We find that our implementation of the convolution in terms of a mixing matrix shows differences at the percent level in comparison to the measurements from a very large set of mock halo catalogs. It is also able to recover unbiased constraints on halo bias parameters in a likelihood analysis of a set of numerical simulations with a total volume of $100\cGpc$. For the level of accuracy required by these tests, the multiplication with the mixing matrix is performed in the time of one second or less.}

\keywords{cosmological parameters from LSS, galaxy clustering, redshift surveys, dark energy experiments}

\begin{document}

\maketitle

%%%%%%%%%%%%%%%%%%%%%%%%%%%%%%%%%%%%%%%%%%%%%%%%%%%%%%%%%%%%%%%%%%
%%%%%%%%%%%%%%%%%%%%%%%%%%%%%%%%%%%%%%%%%%%%%%%%%%%%%%%%%%%%%%%%%%
\section{Introduction}

One particular class of observables to study galaxy clustering is three-dimensional, Fourier-space summary statistics of the galaxy distribution \cite{Peebles1980}. For a Gaussian distribution, all statistical properties are encoded in its power spectrum. However, non-Gaussianity, quantified in non-vanishing, higher-order correlation functions, is induced by gravitational instability, nonlinear bias and redshift-space distortions. The measurement of the galaxy bispectrum, i.e. the three-point function of the galaxy distribution in Fourier space, is therefore expected to complement a power spectrum-only analysis. Such a combined analysis offers specific information on the parameters controlling nonlinear corrections to the power spectrum itself   \cite{Fry1994B, HivonEtal1995, MatarreseVerdeHeavens1997, ScoccimarroEtal1998, Scoccimarro2000B, SefusattiEtal2006, GagraniSamushia2017, YankelevichPorciani2019, DAmicoEtal2020, GualdiVerde2020, HahnEtal2020, HahnVillaescusaNavarro2021, AgarwalEtal2021, IvanovEtal2021A}. 
In addition, the galaxy bispectrum can directly probe a primordial non-Gaussian component, possibly shedding light on the interactions taking place during inflation \cite{ScoccimarroEtal2001B, ScoccimarroSefusattiZaldarriaga2004, SefusattiKomatsu2007, JeongKomatsu2009B, Sefusatti2009, 
TasinatoEtal2014, TellariniEtal2015, MoradinezhadDizgahEtal2018, KaragiannisEtal2018,  MoradinezhadDizgahEtal2021, Barreira2022}. Recent efforts using measurements of the galaxy bispectrum in BOSS are the first steps toward this direction \cite{CabassEtal2022A, DAmicoEtal2022A}.

The analysis of higher-order statistics comes at the price of greater complexity in their measurement, modelling and estimation of their covariance properties. A specific challenge comes from the fact that the conventional FKP-like estimator \cite{ScoccimarroCouchmanFrieman1999, Scoccimarro2000A, Scoccimarro2015} (extending the Feldman-Kaiser-Peacock power spectrum estimator of \cite{FeldmanKaiserPeacock1994}) used in spectroscopic surveys provides as output a non trivial convolution of the underlying galaxy bispectrum with the survey footprint and redshift selection function. The estimator actually computes the three-point function of the Fourier Transform of the product between the galaxy overdensity $\delta(\xv)$ and the window function $W(\xv)$, vanishing outside the observed volume. Consequently, the theoretical model for the bispectrum $B(\kv_1, \kv_2)$ defined as
\begin{equation}
    \langle \delta(\vec{k}_1) \delta(\vec{k}_2) \delta(\vec{k}_3) \rangle = (2\pi)^3 \delta_D(\vec{k}_1 + \vec{k}_2 + \vec{k}_3) B(\vec{k}_1, \vec{k}_2),
\end{equation}
has be to convolved with the bispectrum of the window $B_W(\kv_1, \kv_2) \equiv W(\kv_1) W(\kv_2) W^*(\kv_1 + \kv_2)$ to obtain 
\begin{equation}
\label{bispectrum_window_convolution}
    \tilde{B}(\vec{k}_1, \vec{k}_2) = \int \frac{d^3p_1}{(2\pi)^3} \int \frac{d^3p_2}{(2\pi)^3} B_W(\vec{k}_1-\vec{p}_1, \vec{k}_2 - \vec{p}_2) B(\vec{p}_1, \vec{p}_2)\,,
\end{equation}
that is the quantity to be compared with the measurement. A fast evaluation of this type of integral (at most of the order of few seconds, comparable to a single evaluation of the linear power spectrum by a Boltzmann code), is required for a likelihood analysis, where the matrix multiplication relating $B(\vec{k}_1, \vec{k}_2)$ to $\tilde{B}(\vec{k}_1, \vec{k}_2)$ has to be done for each point sampled in the parameter space.

In the case of the power spectrum, a formulation that turns the convolution integral into a series of one-dimensional Hankel transforms has been proposed by \cite{WilsonEtal2017}. This was later expanded by \cite{BeutlerEtal2017B} to the case of the local plane-parallel estimator of \cite{YamamotoEtal2006, BianchiEtal2015, Scoccimarro2015} and to include wide-angle and general-relativistic effects \cite{CastorinaWhite2018, BeutlerCastorinaZhang2019, CastorinaDiDio2022}. The core technique of this method relies on the FFTLog algorithm \cite{Hamilton2000}, an implementation of the Fast-Fourier Tranform algorithm on logarithmically-spaced points. The Hankel-transform formulation speeds-up the process considerably, as it amounts to evaluating a one-dimensional FFT, scaling as $N\log N$ for $N$ FFT-sample points, instead of the full three-dimensional FFTs required to do power spectrum window convolution via convolution theorem. However, the FFTLog algorithm suffers from the typical FFT problems such as aliasing and ringing, which might give spurious effects if not treated carefully. The formulation in terms of a matrix multiplication, schematically written as $\tilde{P}_\ell(k) = \mathcal{W}_{\ell \ell'} (k,p) P_{\ell'}(p)$ \cite{DAmicoEtal2020, BeutlerMcDonald2021, BeutlerEtal2014A} transfers the FFTLog calculation into the matrix $\mathcal{W}_{\ell \ell'}(k, p)$ that can be pre-computed and tested carefully. Although the matrix multiplication scales, for a single multipole, as $N_k \times N$, where $N_k$ is the number of $k$-modes and $N$ is the number of FFT-sample points, it is still comparable, in complexity, to the Hankel transform formulation since typically $N_k \ll N$.

The bispectrum convolution received so far, understandably, less attention. References \cite{GilMarinEtal2015, GilMarinEtal2015B, GilMarinEtal2017} considered, for a galaxy bispectrum monopole measured according to \cite{Scoccimarro2015},  an approximation that ignores the effect of the convolution of the nonlinear kernel characterizing the tree-level prediction for the bispectrum. This approach works quite well on most triangle shapes, except for squeezed configurations where one side is comparable to the inverse of the characteristic size of the window, $1/R$. This could lead to discarding potentially valuable information on large scales, as we can expect, for instance, in models with local primordial non-Gaussianity \cite{SefusattiCrocceDesjacques2012}.

An alternative is to consider an entirely  different estimator. An example is the tri-polar spherical harmonics (TripoSH) estimator proposed by \cite{SugiyamaEtal2019}, that uses the tensor product of three spherical harmonics as a basis for a bispectrum decomposition. This estimator is the natural extension of the real-space bispectrum estimator for the quantity $B_{\ell}(k_1, k_2)$, function of the two wavevectors $(k_1, k_2)$ and the multipole index $\ell$ characterizing the angle between $\kv_1$ and $\kv_2$. This decomposition has been shown to be computable via two-dimensional Hankel transforms of its three point function multipoles \cite{Szapudi2004}. However, this estimator requires special care with the truncation for the multipoles expansion of bispectrum, since in general this decomposition requires an infinite number of multipoles \cite{SugiyamaEtal2021}.

Another type of estimator recently proposed is the cubic estimator of \cite{Philcox2021}, which bypasses the need of the window convolution of the theoretical modelling altogether. This estimator falls under the class of maximum-likelihood estimators \cite{TegmarkEtal1998, TegmarkTaylorHeavens1997, Hamilton2005a, Hamilton2005b} which, by construction, approach the Cramér-Rao bound on the information content of a given bandpower. This is certainly a very promising way to measure the bispectrum, but it still lacks an analytical understanding of why such estimators, even for the power spectrum, are unbiased in wide-field galaxy surveys, since in principle wide-angle effects are also coupled to the window effects \cite{CastorinaEtal2019, 2021arXiv210613725M}.

Our approach in this work is to provide a theoretical prediction for the redshift-space bispectrum multipoles estimator of \cite{Scoccimarro2015}, which is a natural extension of the FFT-based, variable line-of-sight estimator of the power spectrum estimator \cite{FeldmanKaiserPeacock1994, ScoccimarroEtal1998, BianchiEtal2015, Scoccimarro2015}. Our goal is to write such a convolution as a matrix multiplication, with the mixing matrix computable via a multidimensional Hankel transform.  

This work is organized as follows. A description of the bispectrum estimator that we adopt is presented in section \ref{bisp_estimator}. The bispectrum-window convolution formulation in terms of Hankel transform is laid out in section \ref{sec:window_hankel}. We present the setup of an ideal test in terms of a spherical window in real space in section \ref{sec:set-up} while we show the results of the test in section \ref{sec:results}.  Finally, we present our conclusions in section \ref{conclusion}.

%%%%%%%%%%%%%%%%%%%%%%%%%%%%%%%%%%%%%%%%%%%%%%%%
%%%%%%%%%%%%%%%%%%%%%%%%%%%%%%%%%%%%%%%%%%%%%%%%
\section{Bispectrum multipoles estimator}
\label{bisp_estimator}

We consider the definition of the redshift-space bispectrum multipoles introduced by \cite{ScoccimarroCouchmanFrieman1999} where the orientation of the triangle formed by the wavenumbers $\kv_1$, $\kv_2$, $\kv_3$,  is parameterized in terms of the angle $\theta$ between $\kv_1$ and the line-of-sight and of the angle $\phi$ describing the rotation of $\kv_2$ around $\kv_1$ (see \cite{HashimotoRaseraTaruya2017} for an alternative choice). Expanding this angle-dependence in spherical harmonics $Y_{LM}(\theta,\phi)$, the corresponding multipoles are given by
\be
\label{e:Blm}
B_{LM}(k_1, k_2, k_3)=\int\!\!d\!\cos{\theta} \int\!\!d\phi B(k_1, k_2, k_3, \theta,\phi)\,Y_{LM}(\theta,\phi)\,.
\ee
In practice, we restrict our attention to an expansion in Legendre polynomials ${\mathcal L}_{L}(\cos \theta)$ of the bispectrum averaged over the angle $\phi$, as it is often the case (see \cite{GagraniSamushia2017} for motivation). These are defined as
\begin{align}
\label{e:Bl}
B_L(k_1,k_2,k_3) & = \sqrt{\frac{2L+1}{4\pi}}\,B_{L 0}(k_1,k_2,k_3) \nn\\
& = \frac{2L+1}{4\pi} \int\!\!d\!\cos{\theta} \int d\phi B(k_1, k_2, k_3, \theta,\phi)\,\mathcal{L}_L(\cos \theta)\,.
\end{align}

The estimator proposed by \cite{Scoccimarro2015} can be written as
\be
\label{eq:BLest}
    \hat{B}_{L}(k_1, k_2, k_3) = \frac{2 L+1}{V_B} \int_{k_1}\!\! d^3 q_1\!\! \int_{k_2}\!\! d^3 q_2\!\! \int_{k_3}\!\! d^3 q_3 ~ \delta_D(\vec{q}_{123})  \int_V \frac{d^3x}{V} \,\hat{B}^{\rm local}(\qv_1,\qv_2;\xv) \,{\mathcal L}_{L}(\hat{q}_1\cdot\hat{x})\,,
\ee
where the integration  $\int_{k} d^3 q $ is taken over the spherical shell of radius in the range $k-\Delta k/2 \leq q \leq k+ \Delta k/2$ for the bin width $\Delta k$ and where
\be
V_B(k_1,k_2,k_3) \equiv \int_{k_1}\!\! d^3 q_1\!\! \int_{k_2}\!\! d^3 q_2\!\! \int_{k_3}\!\! d^3 q_3 ~ \delta_D(\vec{q}_{123})\,,
\ee 
adopting the notation for wavevector sums $\qv_{ij}\equiv\qv_i+\qv_j$. The local bispectrum estimator 
\be
\hat{B}^{\rm local}(\qv_1, \qv_2;\xv)  =  \int_V d^3 x_{13} \int_V d^3 x_{23} ~ e^{-i \qv_1 \cdot \xv_{13}} e^{-i \qv_2 \cdot \xv_{23} } \delta(\xv_1) \delta(\xv_2) \delta(\xv_3)
\ee
with $\xv_{ij} = \xv_i-\xv_j$, corresponds to the Fourier Transform of the product $\delta(\xv_1) \delta(\xv_2) \delta(\xv_3)$ where the mean point $\xv=(\xv_1+\xv_2+\xv_3)/3$ provides the direction of the line-of-sight $\hat{n}$, so that $\cos \theta \equiv \hat{q}_1 \cdot \hat{x}$. 

By replacing $\xv\rightarrow \xv_3$, the integrand in eq.~(\ref{eq:BLest}) becomes separable, thus enabling an efficient computation via linear combination of Fast Fourier Transforms. The differences induced in the measurements (and hence in the predictions) between these two definitions of the line-of-sight, the so-called wide-angle effects, are usually negligible in the modelling of the power spectrum multipoles in a typical galaxy survey \cite{CastorinaWhite2018, CastorinaWhite2018b, BeutlerCastorinaZhang2019}\footnote{Note that although wide-angle corrections might be important in the case of a primordial non-Gaussianity analysis where the signal is coming from the large scale, 
it has been neglected in recent power spectrum multipoles analysis of eBOSS quasars, given the high redshift and small observational volume of the sample \cite{CastorinaEtal2019, 2021arXiv210613725M}.}. We will assume the same is true for the bispectrum, although we acknowledge that a quantitative study of wide-angle effects in the bispectrum is missing.

%%%%%%%%%%%%%%%%%%%%%%%%%%%%%%%%%%%%%%%%%%%%%%%%
%%%%%%%%%%%%%%%%%%%%%%%%%%%%%%%%%%%%%%%%%%%%%%%%
\section{Window convolution via Hankel Transform}
\label{sec:window_hankel}

In a cosmological analysis we need to compare the measurements of the estimator in eq.~(\ref{eq:BLest}) to the corresponding theoretical prediction, that is the ensemble average $\tilde{B}_L \equiv \langle \hat{B}_L \rangle$. Replacing $\delta(\xv)\rightarrow \delta(\xv)\,W(\xv)$, with $W(\xv)$ representing here the window as a purely geometrical step function taking values 0 or 1, we obtain
\begin{align}
\label{eq:BL}
    \tilde{B}_L (k_1, k_2, k_3) & =  \frac{2L+1}{V_B} \int_{k_1}\!\! d^3 q_1\!\! \int_{k_2}\!\! d^3 q_2\!\! \int_{k_3}\!\! d^3 q_3\, \delta_D(\vec{q}_{123}) 
    \nn\\ & \times 
    \int_V\!\! \frac{d^3 x_3}{V}\! \int_V\!\! d^3 x_{13}\! \int_V \!\! d^3 x_{23} \, e^{-i \vec{q}_1 \cdot \vec{x}_{13}} e^{-i \vec{q}_2 \cdot \vec{x}_{23}}  \zeta(\vec{x}_{13}, \vec{x}_{23}, \hat{x}_3) 
    \nn \\ & \times 
    W(\xv_1) \, W(\xv_2)\,  W(\xv_3)  \, \mathcal{L}_L (\hat{q}_1 \cdot \hat{x}_3)\,,
\end{align}
where $\zeta(\vec{x}_{13}, \vec{x}_{23}, \hat{x}_3)$ is the galaxy three-point correlation function (3PCF), and its orientation is defined w.r.t. the line-of-sight direction $\hat{x}_3$. 

First, we rewrite the Dirac delta distribution by its integral representation eq.~\eqref{dirac_delta_integral_rep}. Then, by repeated use of the Rayleigh expansion of a plane wave eq.~\eqref{plane_wave_exp}, the addition of spherical harmonics eq.~\eqref{addition_spherical_harmonics}, and the orthogonality of Legendre polynomials eq.~\eqref{LL_ortho}, we can perform the angular integration of the wavevector modes to get
\begin{align}
    \tilde{B}_L (k_1, k_2, k_3) & =
    \frac{8 (4\pi)^2 }{V_B} \int_{k_1} dq_1 q_1^2 \int_{k_2} dq_2 q_2^2 \int_{k_3} dq_3 q_3^2  \int \frac{d^3 x}{V}
    \int_V d^3 x_3 \int_V d^3 x_{13} \int_V d^3 x_{23}   
    \nn \\ & \times 
    \sum_{\substack{\ell, \ell_1, \ell_2 \\ m, m_1, M}} i^{\ell-\ell_1}(2\ell_2+1) j_0(q_3 x) j_{\ell}(q_1 x) j_{\ell_2}(q_2 x) j_{\ell_1}(q_1 x_{13}) j_{\ell_2}(q_2 x_{23}) 
    \nn \\ & \times
    \mathcal{G}_{L \ell_1 \ell}^{M m_1 m} Y_{\ell_1 m_1}^*(\hat{x}_{13}) Y_{\ell m}^*(\hat{x}) Y_{LM}^*(\hat{x}_3) \mathcal{L}_{\ell_2}(\hat{x} \cdot \hat{x}_{23})
    \nn \\ & \times
    \zeta(\vec{x}_{13}, \vec{x}_{23}, \hat{x}_3) ~ W(\xv_1) W(\xv_2) W(\xv_3)\,,
\end{align}
where we have also used the integration of the product of three spherical harmonics eq.~\eqref{integration_3Y} resulting in a Gaunt coefficient, eq.~\eqref{gaunt}. Performing the angular integration of the dummy variable $\hat{x}$ using the addition and orthogonality of spherical harmonics, eq.s~\eqref{addition_spherical_harmonics} and \eqref{YY_ortho}, we arrive at
\begin{align}
\label{B_tilde_rs_1}
\tilde{B}_L (k_1, k_2, k_3) & = 
\frac{8 (4 \pi)^2}{ V_B } \int_{k_1} dq_1 q_1^2 \int_{k_2} dq_2 q_2^2 \int_{k_3} dq_3 q_3^2  \int_V \frac{d^3 x_3}{V} \int_V d^3 x_{13} \int_V d^3 x_{23}   
\nn \\ & \times
\sum_{\substack{\ell_1, \ell_2 \\ M, m_1, m_2}} i^{\ell_2 - \ell_1} I_{\ell_2 \ell_2 0}(q_1, q_2, q_3) j_{\ell_1}(q_1 x_{13}) j_{\ell_2}(q_2 x_{23}) 
\nn \\ & \times 
\mathcal{G}^{M m_1 m_2}_{L \ell_1 \ell_2} Y_{LM}^*(\hat{x}_3) Y_{\ell_1 m_1}^*(\hat{x}_{13}) Y_{\ell_2 m_2}^*(\hat{x}_{23})~ \zeta(\vec{x}_{13}, \vec{x}_{23}, \hat{x}_3) ~ W(\xv_1) W(\xv_2) W(\xv_3)\,,
\end{align}
where we introduce the integral
\begin{equation}\label{I_func}
    I_{\ell_1 \ell_2 \ell_3} (q_1, q_2, q_3) \equiv 4 \pi \int dx x^2 j_{\ell_1}(q_1 x) j_{\ell_2}(q_2 x) j_{\ell_3}(q_3 x).
\end{equation}

The formulation that we seek is a window convolution expressed in terms of a matrix multiplication of the unconvolved bispectrum with a mixing matrix computable via Hankel transforms. Therefore, we need to transform back, and write the 3PCF in terms of the unconvolved, redshift-space bispectrum $B(\pv_1, \pv_2, \hat{x}_3)$. In turn, the latter can be expanded in spherical harmonics as
\begin{equation}
\label{bar_B_definition}
    B(\vec{p}_1, \vec{p}_2, \hat{x}_3) = \sum_{L' M'} \bar{B}_{L' M'}(p_1,p_2,p_3) Y_{L' M'}(\hat{p}_1 \cdot \hat{x}_3, \varphi) \,,
\end{equation}
where now we consider the azimuthal angle  $\varphi$ describing the rotation of $\hat{x}_3$ around $\hat{p}_1$. The two azimuthal coordinates is related by $\varphi=-\phi$ (see also figure~\ref{fig:coord_comparison}). It can be easily shown that the coefficients $\bar{B}_{LM}$ of this expansion are related to those defined in  eq.~\eqref{e:Blm} via $\bar{B}_{LM} = (-1)^M B_{L, -M}$ so that
\begin{equation}
\label{B_in_BLM}
    B(\vec{p}_1, \vec{p}_2, \hat{x}_3) =\sum_{L' M'} B_{L', -M'}(p_1,p_2,p_3) (-1)^{M'} Y_{L' M'}(\hat{p}_1 \cdot \hat{x}_3, \varphi)\,.
\end{equation}
Note that with a fixed $\hat{x}_3$, in this new coordinate definition $Y_{LM}(\hat{p}_1 \cdot \hat{x}_3, \varphi) = Y_{LM}(\hat{p}_1$), which is completely specified by the orientation of a single unit vector $\hat{p}_1$ with respect to $\hat{x}_3$, without the need to mention the orientation of $\hat{p}_2$ as in eq.~\eqref{e:Blm}. We can therefore write 
\begin{align}
\label{3pcf_in_terms_of_bp1p2p3}
    \zeta(\vec{x}_{13}, \vec{x}_{23}, \hat{x}_3)  = & 
    \int\! \frac{d^3 p_1}{(2\pi)^3} \int\! \frac{d^3 p_2}{(2\pi)^3} \int\! \frac{d^3 p_3}{(2\pi)^3} e^{i \pv_1 \cdot \xv_{13}} e^{i \pv_2 \cdot \xv_{23}} B(p_1, p_2, p_3, \hat{x}_3) (2\pi)^3 \delta_D(\vec{p}_{123})
    \nn \\ = & 
    \int\! \frac{d^3 p_1}{(2\pi)^3} \int\! \frac{d^3 p_2}{(2\pi)^3} \int\! \frac{d p_3}{2\pi^2} p_3^2~ e^{i \vec{p}_1 \cdot \vec{x}_{13}} e^{i \vec{p}_2 \cdot \vec{x}_{23}} B(p_1, p_2, p_3, \hat{x}_3) 
    \nn \\ & \times 
    \sum_{\ell} 4\pi (-1)^\ell (2\ell+1) \int dx ~ x^2  j_\ell(p_1 x) j_\ell(p_2 x) j_0(p_3 x) \mathcal{L}_\ell(\hat{p}_1 \cdot \hat{p}_2) 
    \nn \\ = & 
    \int\! \frac{d p_1}{2 \pi^2} p_1^2 \int\! \frac{d p_2}{2 \pi^2} p_2^2 \int\! \frac{d p_3}{2 \pi^2} p_3^2 \sum_{L' M'} B_{L',-M'}(p_1, p_2, p_3) (-1)^{M'} 
    \nn \\ & \times 
    \sum_{\ell} I_{\ell \ell 0}(p_1, p_2, p_3) \sum_{m  m' \ell'} 4\pi  i^{\ell' - \ell} ~\mathcal{G}_{L' \ell \ell'}^{M' m m'}
    \nn \\ & \times  
    j_{\ell'}(p_1 x_{13}) j_\ell(p_2 x_{23}) Y^*_{\ell' m'}(\hat{x}_{13}) Y^*_{\ell m}(\hat{x}_{23})\,,
\end{align}
where we have used the addition of spherical harmonics identity eq.~\eqref{addition_spherical_harmonics}, the contraction of spherical harmonics eq.~\eqref{contraction_Y} and the integration of spherical harmonics along solid angle eq.~\eqref{integrate_Y_ellm} to simplify the integral. Plugging back eq.~\eqref{3pcf_in_terms_of_bp1p2p3} into eq.~\eqref{B_tilde_rs_1} gives us our final expression for window convolution of the bispectrum as a matrix multiplication
\begin{align}
\label{eq:BconvLM}
\tilde{B}_L(k_1, k_2, k_3) = & \int \frac{d p_1}{2 \pi^2}p_1^2 \int \frac{d p_2}{2 \pi^2}p_2^2 \int \frac{d p_3}{2 \pi^2}p_3^2 ~ \sum_{L' M'} B_{L' M'}(p_1, p_2, p_3) 
\nn \\ & \times 
\sum_{\ell} I_{\ell \ell 0}(p_1, p_2, p_3) \mathcal{Q}_{L' ,-M', \ell}^{L}(k_1, k_2, k_3; p_1, p_2)\,,
\end{align}
where the matrix elements are intended to be the triplets $\{p_1,p_2,p_3\}$. The mixing matrix is given in terms of a product where the first factor is the $I_{\ell \ell 0}(p_1, p_2, p_3)$ function of eq.~\eqref{I_func} that enforces the triangle condition on $\{p_1, p_2, p_3\}$ \cite{MehremLonderganMacfarlane1991}. This can be written as 
\begin{equation}
\label{I_ell_ell_0}
    I_{\ell \ell 0}(x, y, z) = (-1)^\ell \frac{\pi^2}{xyz} \vartheta(|\hat{x} \cdot \hat{y}|) \mathcal{L}_\ell\Big(\frac{z^2 - x^2 - y^2}{2xy}\Big)\,,
\end{equation}
where $\vartheta(\Delta)$ is the (modified) Heaviside step function that takes value of 1 for $\Delta < 1$, 1/2 for $\Delta = 1$ while vanishing everywhere else. The window function contribution is accounted for by the second factor, defined as follows \begin{align} 
\label{Q_L_redshift}
    \mathcal{Q}^{L}_{L' M' \ell}(k_1, k_2, k_3; p_1, p_2) \equiv &
    \sum_{\ell_1, \ell_2, \ell'} \frac{128 \pi^2}{V_B} \int_{k_1} dq_1 q_1^2 \int_{k_2} dq_2 q_2^2 \int_{k_3} dq_3 q_3^2 ~ I_{\ell_2 \ell_2 0}(q_1, q_2, q_3) 
    \nn \\ & \times 
    \mathcal{W}^L_{L' M' \ell \ell' \ell_1 \ell_2}(q_1, q_2; p_1, p_2)\,,
\end{align}
where we introduce the function
\begin{align}
\label{two_dim_hankel_W}
\mathcal{W}^{L}_{L' M' \ell \ell' \ell_1 \ell_2}(q_1, q_2; p_1, p_2) \equiv &
\,(4\pi)^2 \int\! d x_{13}\, x_{13}^2 \int\! d x_{23}\, x_{23}^2 \, j_{\ell'}(p_1 x_{13}) \, j_{\ell} (p_2 x_{23})
\nn \\ & \times 
\Big [j_{\ell_1}(q_1 x_{13}) \, j_{\ell_2} (q_2 x_{23}) \, Q^{L}_{L' M' \ell \ell' \ell_1 \ell_2}(x_{13}, x_{23}) \Big]\,,
\end{align}
that can be evaluated as a two-dimensional Hankel Transform
of the product of two spherical Bessel functions with the 3PCF multipoles of the window function defined as follows
\begin{align}
\label{3PCF_multipoles_redshifts}
Q^{L}_{L' M'  \ell \ell' \ell_1 \ell_2 }(x_{13}, x_{23}) \equiv &  (-1)^{M'} \sum_{\tilde{\ell}_1, \tilde{\ell}_2} \sum_{\substack{M, m_1, m_2\\ m, m', \tilde{m}_1, \tilde{m}_2}} 4\pi i^{\ell' - \ell + \ell_2 - \ell_1} \, \mathcal{G}_{L \ell_1 \ell_2 }^{M m_1 m_2} \, \mathcal{G}_{L' \ell \ell'}^{M' m m'} \, \mathcal{G}_{\ell_1 \ell' \tilde{\ell}_1 }^{m_1 m' \tilde{m}_1} \, \mathcal{G}_{\ell_2 \ell \tilde{\ell}_2 }^{m_2 m \tilde{m}_2} \,
\nn \\ & \times
\int_V \frac{d^3 x_3}{V} \int \frac{d^2 \hat{x}_{13}}{4\pi} \int \frac{d^2 \hat{x}_{23}}{4\pi} Y_{LM}^*(\hat{x}_3) Y_{\tilde{\ell}_1 \tilde{m}_1}(\hat{x}_{13}) Y_{\tilde{\ell}_2 \tilde{m}_2}(\hat{x}_{23})
\nn \\ & \times
W(\vec{x}_3 + \vec{x}_{13}) W(\vec{x}_3 + \vec{x}_{23}) W(\vec{x}_3),
\end{align}
where we have used eq.~\eqref{contraction_Y} to simplify the expression.
We will leave to future work the investigation of the best method to measure the window 3PCF multipoles, but see appendix~\ref{app:window3PCF} for some possible directions. 

So far we derived our results including the full integrals over the wavenumbers bins characterising the estimator of eq.~\eqref{eq:BLest}. For $k\gg\Delta k$ we can consider a simplified expression in the thin-shell approximation (we estimate in section \ref{binning_effect} the errors associated with this choice). This amounts to assume the integrand a slowly-varying function of $\{q_1,q_2,q_3\}$ so that
\begin{equation}\label{thin_shell_approx}
    \int_{k_1}\! dq_1\, q_1^2 \int_{k_2}\! dq_2\, q_2^2 \int_{k_3}\! dq_3\, q_3^2 ~F(q_1, q_2, q_3) \simeq F(k_1, k_2, k_3)\,\int_{k_1}\! dq_1\, q_1^2 \int_{k_2}\! dq_2\, q_2^2 \int_{k_3}\! dq_3\, q_3^2\,,
\end{equation}
simplifying eq.~\eqref{Q_L_redshift} into 
\be
\label{Q_L_redshift_final} 
    \mathcal{Q}^{L}_{L' M' \ell}(k_1, k_2, k_3; p_1, p_2) \simeq \sum_{\ell_1,\ell_2, \ell'} 16 \pi^2 \frac{I_{\ell_2 \ell_2 0}(k_1, k_2, k_3)}{I_{000}(k_1, k_2, k_3)} \mathcal{W}^L_{L' M' \ell \ell' \ell_1 \ell_2}(k_1, k_2; p_1, p_2)\,.
\ee
\tikzset{every picture/.style={line width=0.75pt}} %set default line width to 0.75pt

\begin{figure}[t!]
\begin{tikzpicture}[x=0.75pt,y=0.75pt,yscale=-1,xscale=1] 
%uncomment if require: \path (0,258); %set diagram left start at 0, and has height of 258

%Straight Lines [id:da7186609983545735] 
\draw    (166,152) -- (164.69,39.5) ;
\draw [shift={(164.67,37.5)}, rotate = 449.33] [color={rgb, 255:red, 0; green, 0; blue, 0 }  ][line width=0.75]    (10.93,-4.9) .. controls (6.95,-2.3) and (3.31,-0.67) .. (0,0) .. controls (3.31,0.67) and (6.95,2.3) .. (10.93,4.9)   ;
%Straight Lines [id:da9520950032573321] 
\draw    (166,152) -- (67.27,225.31) ;
\draw [shift={(65.67,226.5)}, rotate = 323.40999999999997] [color={rgb, 255:red, 0; green, 0; blue, 0 }  ][line width=0.75]    (10.93,-4.9) .. controls (6.95,-2.3) and (3.31,-0.67) .. (0,0) .. controls (3.31,0.67) and (6.95,2.3) .. (10.93,4.9)   ;
%Straight Lines [id:da43038410070380073] 
\draw    (166,152) -- (302.67,153.48) ;
\draw [shift={(304.67,153.5)}, rotate = 180.62] [color={rgb, 255:red, 0; green, 0; blue, 0 }  ][line width=0.75]    (10.93,-4.9) .. controls (6.95,-2.3) and (3.31,-0.67) .. (0,0) .. controls (3.31,0.67) and (6.95,2.3) .. (10.93,4.9)   ;
%Straight Lines [id:da23084698783770574] 
\draw    (166,152) -- (164.71,69.67) ;
\draw [shift={(164.67,66.67)}, rotate = 449.1] [fill={rgb, 255:red, 0; green, 0; blue, 0 }  ][line width=0.08]  [draw opacity=0] (10.72,-5.15) -- (0,0) -- (10.72,5.15) -- (7.12,0) -- cycle    ;
%Straight Lines [id:da6300113747731892] 
\draw    (166,152) -- (106.09,108.43) ;
\draw [shift={(103.67,106.67)}, rotate = 396.03] [fill={rgb, 255:red, 0; green, 0; blue, 0 }  ][line width=0.08]  [draw opacity=0] (10.72,-5.15) -- (0,0) -- (10.72,5.15) -- (7.12,0) -- cycle    ;
%Straight Lines [id:da4726316391139328] 
\draw    (166,152) -- (227.7,80.93) ;
\draw [shift={(229.67,78.67)}, rotate = 490.96] [fill={rgb, 255:red, 0; green, 0; blue, 0 }  ][line width=0.08]  [draw opacity=0] (10.72,-5.15) -- (0,0) -- (10.72,5.15) -- (7.12,0) -- cycle    ;
%Shape: Arc [id:dp5998127440127706] 
\draw  [draw opacity=0] (122.49,119.74) .. controls (124.22,115.65) and (126.9,111.87) .. (130.49,108.77) .. controls (139.73,100.76) and (152.48,99.41) .. (162.88,104.29) -- (150.12,131.45) -- cycle ; \draw   (122.49,119.74) .. controls (124.22,115.65) and (126.9,111.87) .. (130.49,108.77) .. controls (139.73,100.76) and (152.48,99.41) .. (162.88,104.29) ;
%Straight Lines [id:da43663544647649855] 
\draw  [dash pattern={on 0.84pt off 2.51pt}]  (225.67,86.67) -- (223,203) ;
%Straight Lines [id:da9196813058060049] 
\draw  [dash pattern={on 0.84pt off 2.51pt}]  (166,152) -- (223,203) ;
%Shape: Arc [id:dp323552989546974] 
\draw  [draw opacity=0] (189.16,173.34) .. controls (183.83,181.53) and (174.61,186.96) .. (164.11,187) .. controls (153.69,187.04) and (144.49,181.76) .. (139.09,173.72) -- (164,157) -- cycle ; \draw   (189.16,173.34) .. controls (183.83,181.53) and (174.61,186.96) .. (164.11,187) .. controls (153.69,187.04) and (144.49,181.76) .. (139.09,173.72) ;
%Straight Lines [id:da8970902456086142] 
\draw    (472.97,157.08) -- (471.66,44.58) ;
\draw [shift={(471.63,42.58)}, rotate = 449.33] [color={rgb, 255:red, 0; green, 0; blue, 0 }  ][line width=0.75]    (10.93,-4.9) .. controls (6.95,-2.3) and (3.31,-0.67) .. (0,0) .. controls (3.31,0.67) and (6.95,2.3) .. (10.93,4.9)   ;
%Straight Lines [id:da6578661845342204] 
\draw    (472.97,157.08) -- (374.24,230.39) ;
\draw [shift={(372.63,231.58)}, rotate = 323.40999999999997] [color={rgb, 255:red, 0; green, 0; blue, 0 }  ][line width=0.75]    (10.93,-4.9) .. controls (6.95,-2.3) and (3.31,-0.67) .. (0,0) .. controls (3.31,0.67) and (6.95,2.3) .. (10.93,4.9)   ;
%Straight Lines [id:da6739476356930632] 
\draw    (472.97,157.08) -- (609.63,158.56) ;
\draw [shift={(611.63,158.58)}, rotate = 180.62] [color={rgb, 255:red, 0; green, 0; blue, 0 }  ][line width=0.75]    (10.93,-4.9) .. controls (6.95,-2.3) and (3.31,-0.67) .. (0,0) .. controls (3.31,0.67) and (6.95,2.3) .. (10.93,4.9)   ;
%Straight Lines [id:da5484387925140995] 
\draw    (472.97,157.08) -- (471.68,74.75) ;
\draw [shift={(471.63,71.75)}, rotate = 449.1] [fill={rgb, 255:red, 0; green, 0; blue, 0 }  ][line width=0.08]  [draw opacity=0] (10.72,-5.15) -- (0,0) -- (10.72,5.15) -- (7.12,0) -- cycle    ;
%Straight Lines [id:da39854664024581576] 
\draw    (472.97,157.08) -- (413.06,113.51) ;
\draw [shift={(410.63,111.75)}, rotate = 396.03] [fill={rgb, 255:red, 0; green, 0; blue, 0 }  ][line width=0.08]  [draw opacity=0] (10.72,-5.15) -- (0,0) -- (10.72,5.15) -- (7.12,0) -- cycle    ;
%Straight Lines [id:da05111202231552492] 
\draw    (472.97,157.08) -- (534.67,86.02) ;
\draw [shift={(536.63,83.75)}, rotate = 490.96] [fill={rgb, 255:red, 0; green, 0; blue, 0 }  ][line width=0.08]  [draw opacity=0] (10.72,-5.15) -- (0,0) -- (10.72,5.15) -- (7.12,0) -- cycle    ;
%Shape: Arc [id:dp9943663161324999] 
\draw  [draw opacity=0] (471.75,116.42) .. controls (473.52,114.98) and (475.54,113.84) .. (477.78,113.05) .. controls (487.3,109.72) and (497.87,113.98) .. (504.3,122.83) -- (486.64,138.37) -- cycle ; \draw   (471.75,116.42) .. controls (473.52,114.98) and (475.54,113.84) .. (477.78,113.05) .. controls (487.3,109.72) and (497.87,113.98) .. (504.3,122.83) ;
%Straight Lines [id:da494650894788323] 
\draw  [dash pattern={on 0.84pt off 2.51pt}]  (532.63,91.75) -- (529.97,208.08) ;
%Straight Lines [id:da7059272254311852] 
\draw  [dash pattern={on 0.84pt off 2.51pt}]  (472.97,157.08) -- (529.97,208.08) ;
%Shape: Arc [id:dp6632728789383021] 
\draw  [draw opacity=0] (496.13,178.42) .. controls (490.8,186.61) and (481.58,192.04) .. (471.08,192.08) .. controls (460.66,192.12) and (451.46,186.84) .. (446.05,178.8) -- (470.97,162.08) -- cycle ; \draw   (496.13,178.42) .. controls (490.8,186.61) and (481.58,192.04) .. (471.08,192.08) .. controls (460.66,192.12) and (451.46,186.84) .. (446.05,178.8) ;

% Text Node
\draw (65.33,230.83) node   [align=left] {\begin{minipage}[lt]{15.413356pt}\setlength\topsep{0pt}
$\displaystyle \hat{x}$
\end{minipage}};
% Text Node
\draw (318.33,149.83) node   [align=left] {\begin{minipage}[lt]{15.413356pt}\setlength\topsep{0pt}
$\displaystyle \hat{y}$
\end{minipage}};
% Text Node
\draw (168.33,22.25) node   [align=left] {\begin{minipage}[lt]{15.413356pt}\setlength\topsep{0pt}
$\displaystyle \hat{z}$
\end{minipage}};
% Text Node
\draw (170,61.4) node [anchor=north west][inner sep=0.75pt]  [font=\footnotesize]  {$\hat{p}_{1}$};
% Text Node
\draw (102,90.4) node [anchor=north west][inner sep=0.75pt]  [font=\footnotesize]  {$\hat{x}_3$};
% Text Node
\draw (233,73.4) node [anchor=north west][inner sep=0.75pt]  [font=\footnotesize]  {$\hat{p}_{2}$};
% Text Node
\draw (143,116.4) node [anchor=north west][inner sep=0.75pt]  [font=\footnotesize]  {$\theta $};
% Text Node
\draw (157,163.4) node [anchor=north west][inner sep=0.75pt]  [font=\footnotesize]  {${\phi}$};
% Text Node
\draw (372.3,236.33) node   [align=left] {\begin{minipage}[lt]{15.413356pt}\setlength\topsep{0pt}
$\displaystyle \hat{x}'$
\end{minipage}};
% Text Node
\draw (625.3,156.33) node   [align=left] {\begin{minipage}[lt]{15.413356pt}\setlength\topsep{0pt}
$\displaystyle \hat{y}' $
\end{minipage}};
% Text Node
\draw (476.97,66.48) node [anchor=north west][inner sep=0.75pt]  [font=\footnotesize]  {$\hat{p}_{1}$};
% Text Node
\draw (405.97,89.48) node [anchor=north west][inner sep=0.75pt]  [font=\footnotesize]  {$\hat{p}_{2}$};
% Text Node
\draw (539.97,78.48) node [anchor=north west][inner sep=0.75pt]  [font=\footnotesize]  {$\hat{x}_3$};
% Text Node
\draw (478.85,120.77) node [anchor=north west][inner sep=0.75pt]  [font=\footnotesize]  {$\theta $};
% Text Node
\draw (464.97,169.48) node [anchor=north west][inner sep=0.75pt]  [font=\footnotesize]  {$\varphi $};
% Text Node
\draw (476.33,26.25) node   [align=left] {\begin{minipage}[lt]{15.413356pt}\setlength\topsep{0pt}
$\displaystyle \hat{z}' $
\end{minipage}};
% Text Node
\draw (158,216) node [anchor=north west][inner sep=0.75pt]   [align=left] {($a$)};
% Text Node
\draw (468,219.33) node [anchor=north west][inner sep=0.75pt]   [align=left] {($b$)};
\end{tikzpicture}
\caption{Comparison of (a) Scoccimaro's coordinate definition used in  eq.~\eqref{e:Blm} with (b) coordinate used to define the bispectrum multipoles in eq.~\eqref{bar_B_definition}. The relation between the two azimuthal coordinates definition is $\varphi=-\phi$, which can be proven by taking $\hat{z}' = \hat{z}$, $\hat{x}' = \cos \phi~ \hat{x} + \sin \phi~ \hat{y}$  and $\hat{y}' = -\sin \phi~ \hat{x} + \cos \phi ~\hat{y}$.}
\label{fig:coord_comparison}
\end{figure}

%%%%%%%%%%%%%%%%%%%%%%%%%%%%%%%%%%%%%%%%%%%%%%%%%%%%%%%%%%%%%%%
%%%%%%%%%%%%%%%%%%%%%%%%%%%%%%%%%%%%%%%%%%%%%%%%%%%%%%%%%%%%%%%
\section{Ideal test set-up}
\label{sec:set-up}

In the previous section, we have laid out an expression that works for general window functions in redshift space. Here we want to  test the performance of our formulation. We work with measurements of the bispectrum monopole in real space in the ideal case of a spherical window applied to a distribution of particles (halos) of constant density. This allows us to work out an analytical description of the window and thereby to disentangle possible systematic and numerical errors related to the convolution and its implementation, from those affecting the estimation of the window 3PCF from a random catalog. We leave the application to more realistic survey geometries in redshift space to future work.

%%%%%%%%%%%%%%%%%%%%%%%%%%%%%%%%%%%%%%%%%%%%%%%%%%%%%%%%%%%%%%%
\subsection{Data}

We test our model on the Minerva set of N-body simulations (first presented in \cite{GriebEtal2016}), comprising 298 realizations each evolving 1000${}^3$ dark-matter particles in a box of a side 1500$\Mpc$ assuming a flat $\Lambda$CDM cosmology with $h=0.695$, $\Omega_m$ = 0.285, $\Omega_b$ = 0.046, $n_s = 0.9632$, $\sigma_8 = 0.828$.  In particular, we focus on the halo distribution at $z=1$, taking advantage of several measurements and results already obtained in \cite{OddoEtal2020} and \cite{OddoEtal2021}.

Alongside the N-body simulations, we make use of a much larger set of 10,000 realisations of mock halo catalogs obtained with the \pin{} code \cite{MonacoTheunsTaffoni2002, MonacoEtal2013, MunariEtal2017} based on third-order Lagrangian perturbation theory. The mocks are run on the same box and assume the same cosmology and their mass threshold is defined in order to match the amplitude of the Minerva halo power spectrum at large-scales, including shot-noise (see \cite{OddoEtal2020} for further details). This very large set of mocks serves two purposes. On one hand it provides a robust estimate of the bispectrum covariance matrix. On the other hand, it offers an illustration of the window effects on the average of bispectrum measurements where statistical errors are severely suppressed, below the level of the systematic errors under study.

%%%%%%%%%%%%%%%%%%%%%%%%%%%%%%%%%%%%%%%%%%%%%%%%%%%%%%%%%%%%%%%
\subsection{Measurements}

All the bispectrum measurements performed on the full box with periodic boundary conditions take advantage of the estimator already introduced in eq.~\eqref{eq:BLest}. Its implementation follows the efficient approach of \cite{Scoccimarro2015} based on the FFT of the halo density obtained by fourth-order interpolation on a grid and the interlacing technique to reduce aliasing \cite{SefusattiEtal2016}.

The measurements accounting for the window function adopt the same estimator but with the density field $\delta(\xv)$ replaced by the auxiliary field
\be
F(\xv)=\frac{1}{\bar{n}}\left[n(\xv)-\alpha n_r(\xv)\right]\,,
\ee
where $n(\xv)$ is the density of the halo catalog, with mean $\bar{n}$, while $n_r(\xv)$ is the density of a random catalogs of points with no spatial correlation. The constant $\alpha$ represents the inverse of the ratio between the total number of random points to the total number of halos in the given realisation, a number close to 50. The estimator for the bispectrum of $F$ corresponds therefore to the usual FKP estimator in the simplified situation of a constant density, which implies FKP weights are also constant and can therefore be ignored. For these  measurements both the halo and the random catalogs are distributed over a sphere of volume $700^3\cMpc$ extracted from the same realisations used for the full-box measurements, producing $298$ and $10,000$ bispectra respectively for the Minerva simulations and the \pin{} mocks. The enclosing box where the FT is defined for the window measurement is the same as the simulation box, that is of side $1500\Mpc$.

We consider all triangular configurations that one can define from wavenumber bins of size $\Delta k=2 k_f$, $k_f\equiv 2\pi/L$ being the fundamental frequency of the box of size $L$, starting with the first bin centered at $1.5\,k_f$. This results in a total of 360 triangles up to $k_\mathrm{max} \approx 0.12\kMpc$, ordered such that $k_1 \geq k_2 \geq k_3$ to avoid redundancy. We include the so-called ``open triangles'', that is those configurations where the bin centers $\{k_1,k_2,k_3\}$ satisfy the inequalities $0<k_1 - k_2 - k_3 \leq 1.5 \Delta k$ which allow the fundamental triangle $\{\qv_1,\qv_2,\qv_3\}$, with $k_i-\Delta k/2\leq |\qv_i|<k_i+\Delta k/2$ to satisfy the triangle condition $q_1 - q_2 - q_3 \leq 0$. The $360$ triangles exclude instead all configurations involving a side smaller than the {\em effective} fundamental frequency of the spherical window, defined as $k_f^{\rm sphere}=2\pi/700\kMpc\simeq 2.14 k_f$. These wavenumbers are in fact not relevant as they correspond to wavelengths larger than the effective size of the sphere.

\begin{figure}[t!]
    \includegraphics[width=0.49\textwidth]{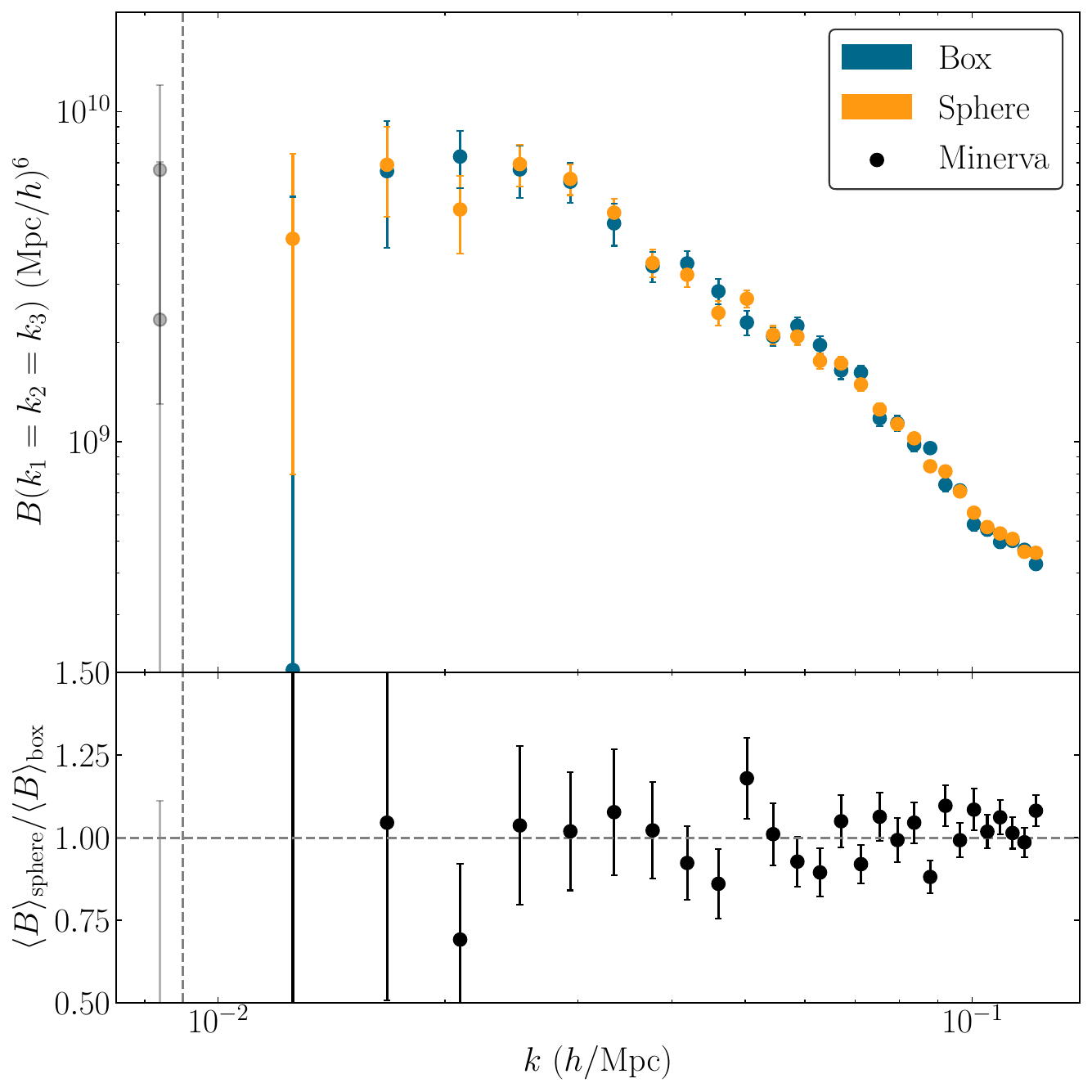}
    \includegraphics[width=0.49\textwidth]{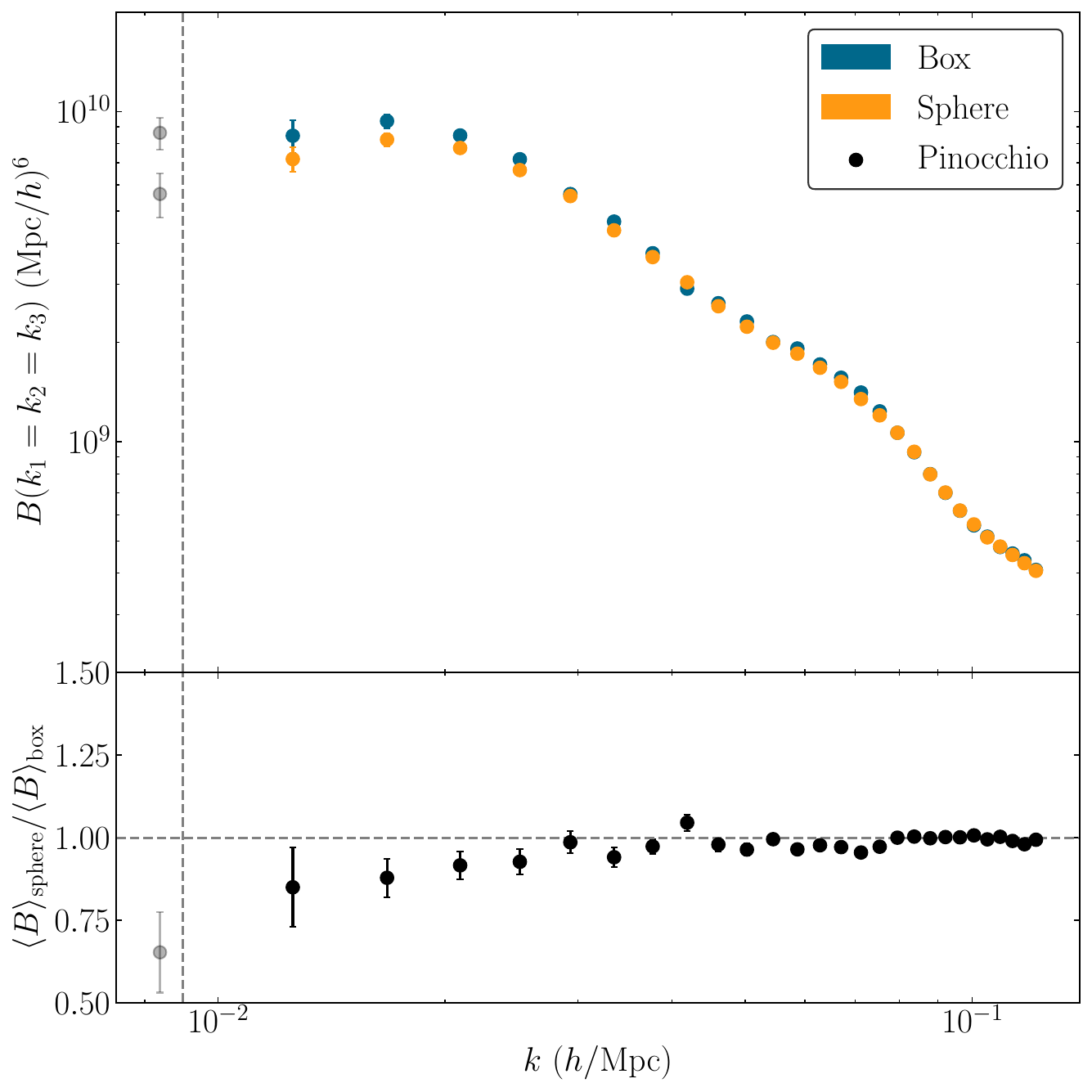}
    \centering
    \caption{\emph{Top panels:} mean of the bispectrum equilateral configurations, $\hat{B}_0(k,k,k)$, as a function of $k$ for both the box (blue) and spherical window (orange) measurements and for both the Minerva simulations ({\em left}) and \pin{} box ({\em right}). \emph{Bottom panels:} ratio of the sphere to box measurements for simulations and mocks. The vertical dashed line indicates the effective fundamental frequency of the sphere $k_f^\mathrm{sphere} = 2\pi/V_{\rm sphere}^{1/3}$. The typical suppression induced by the window on large scales is only detectable in the case of the very large set of \pin{} mocks.}
    \label{fig:min_pin_eq}
\end{figure}

The top panels of figure~\ref{fig:min_pin_eq} show the mean of the measured bispectrum equilateral configurations, $\hat{B}_0(k,k,k)$, as a function of $k$ for both the box (blue) and spherical window (orange)  and for both the Minerva simulations (left) and \pin{} box (right). The bottom panels show the ratio of the sphere to box measurements for simulations and mocks. The typical suppression induced by the window on large scales is only detectable in the case of the very large set of mocks: for the even relatively large set of simulations ($\sim 1000\cGpc$ of total volume), the statistical errors are larger than any window effect. We stress, in this respect, that in the bispectrum case, where the signal is distributed over a large number of configurations, a systematic effect equally distributed over all configurations can have a sizable effect, even if small when compared to the statistical error for the single triangle.

In what follows we refer to the bispectrum monopole measurements $\hat{B}_0$ simply as $\hat{B}$, dropping the subscript for simplicity.

%%%%%%%%%%%%%%%%%%%%%%%%%%%%%%%%%%%%%%%%%%%%%%%%%%%%%%%%%%%%%%%
\subsection{Theoretical model}

The theoretical model that we adopt for the real-space bispectrum is the usual tree-level model in Standard Perturbation Theory (SPT), see e.g. \cite{BernardeauEtal2002} and \cite{ChanScoccimarroSheth2012, BaldaufEtal2012} for the inclusion of a term due to nonlocal bias. This is the same model already employed in \cite{OddoEtal2020} to whom we refer the reader for its detailed description and its fit to the same box bispectrum measurements considered here. We briefly summarise the full expression, given by
\begin{equation}\label{full_model}
    B(\pv_1, \pv_2) = b_1^3 B_{\mathrm{TL}}(\pv_1, \pv_2) + b_2 b_1^2 \Sigma(\pv_1, \pv_2) + 2b_{\mathcal{G}_2} b_1^2 K(\pv_1, \pv_2) + B_\mathrm{SN}(\pv_1, \pv_2)\,,
\end{equation}
where 
\begin{equation}
    B_{\mathrm{TL}}(\vec{p}_1, \vec{p}_2) = 2 F_2(\vec{p}_1, \vec{p}_2)P_L(p_1) P_L(p_2) + \mathrm{2~perms.}\,,
\end{equation}
represents the tree-level matter bispectrum, $F_2$ being the standard quadratic PT kernel and $P_L(p)$ the linear matter power spectrum, while the quadratic local and nonlocal bias contributions are given by
\begin{align}
\label{spt_kernels}
\Sigma(\pv_1, \pv_2) &= P_L(p_1) P_L(p_2) + \mathrm{2 ~ perms.} \\
K(\vec{p}_1, \vec{p}_2) &= \left[\left(\hat{p}_1 \cdot \hat{p}_2 \right)^2 - 1\right] P_L(p_1) P_L(p_2) + \mathrm{2 ~ perms.}\,.
\end{align}
Finally, the shot-noise contribution is 
\begin{equation}
\label{b_SN}
    B_{\mathrm{SN}}(\vec{p}_1, \vec{p}_2) = \frac{1}{\bar{n}}(1+\alpha_1)b_1^2\left[P_L(p_1) + P_L(p_2) + P_L(p_3)\right] + \frac{1}{\bar{n}^2}(1+\alpha_2)\,,
\end{equation}
with the $\alpha_1$ and $\alpha_2$ parameters allowing deviations from the Poisson prediction. We assume all cosmological parameters are known, thus the model depends only on five, free bias and shot-noise parameters: $b_1$, $b_2$, $b_{\mathcal{G}_2}$, $\alpha_1$ and $\alpha_2$.

%%%%%%%%%%%%%%%%%%%%%%%%%%%%%%%%%%%%%%%%%%%%%%%%%%%%%%%%%%%%%%%
\subsection{Window convolution}
\label{sec:real-space-window}

Reducing eq.~\eqref{eq:BconvLM} to the case of the bispectrum monopole, i.e. $L=0$, and in real-space, $L'=0$, $M'=0$ in eq.~\eqref{3PCF_multipoles_redshifts}, to describe the measurements of our test, gives the following expression for $\tilde{B}=\tilde{B}_0$ (see section \ref{sec:real-space-window-derivation} for the derivation)
\be
\label{2dfftlog-approach}
\tilde{B}(q_1, q_2, q_3) = \int \frac{d p_1}{2 \pi^2} p_1^2 \int \frac{d p_2}{2 \pi^2} p_2^2 \int \frac{d p_3}{2 \pi^2} p_3^2 ~B(p_1, p_2, p_3) \sum_{\ell} I_{\ell \ell 0}(p_1, p_2, p_3) \mathcal{Q}_\ell(q_1, q_2, q_3;p_1, p_2),
\ee
with
\be
\label{Qell_real_space}
    \mathcal{Q}_\ell(q_1, q_2, q_3;p_1, p_2) \equiv \sum_{\ell', \ell''}    
    (2\ell''+1) (-1)^{\ell''}\begin{pmatrix}
    \ell & \ell' & \ell'' \\
    0 & 0 & 0
    \end{pmatrix}^2
    \mathcal{W}_{\ell \ell' \ell''}(q_1, q_2;p_1, p_2),
\ee
and
\begin{align}
\label{W_llplpp}
\mathcal{W}_{\ell \ell' \ell''}(q_1, q_2;p_1, p_2) \equiv & (4\pi)^2 (2\ell+1)\int d x_{13} x_{13}^2 \int d x_{23} x_{23}^2 ~ j_{\ell}(p_1 x_{13}) j_{\ell}(p_2 x_{23}) 
\nn \\ & \times 
\Big[  \mathcal{L}_{\ell''}(\hat{q}_1 \cdot \hat{q}_2)j_{\ell''}(q_1 x_{13}) j_{\ell''}(q_2 x_{23}) Q_{\ell'}(x_{13}, x_{23}) \Big],  
\end{align}
where $Q_\ell(x_{13}, x_{23})$ is the 3PCF multipoles of the window function
\begin{equation}
Q_\ell(x_{13}, x_{23}) \equiv (2\ell +1) \int \frac{d^2 \hat{x}_{13}}{4 \pi} \int \frac{d^2 \hat{x}_{23}}{4 \pi} Q(\vec{x}_{13}, \vec{x}_{23}) \mathcal{L}_\ell(\hat{x}_{13} \cdot \hat{x}_{23}),
\end{equation}
with
\begin{equation}\label{Q_3pcf}
    Q(\vec{x}_{13}, \vec{x}_{23}) \equiv \int_V \frac{d^3 x_{3}}{V}~ W(\vec{x}_{3}) W(\vec{x}_{3} + \vec{x}_{13}) W(\vec{x}_{3} + \vec{x}_{23}).
\end{equation}
Note that to derive eq.~\eqref{Qell_real_space}, we have used the thin-shell approximation eq.~\eqref{thin_shell_approx} and eq.~\eqref{I_ell_ell_0}.

In addition to our approach to the convolution we also consider, for comparison, the approximation adopted in the analysis of \cite{GilMarinEtal2015, GilMarinEtal2015B, GilMarinEtal2017} (see also \cite{DAmicoEtal2020, DAmicoEtal2022A} for more recent work), which ignores the effect of the window convolution on the nonlinear kernel of the tree-level prediction while describing it in terms of a convolution of the power spectra. Symbolically we have
\be
\label{1dfftlog-approx}
 \tilde{B}[P_L]  \simeq B[\tilde{P}_L]\,,
\ee
where $\tilde{P}_L(k)$ is the window convolution of the linear power spectrum, evaluated by the usual one-dimensional FFTLog method \cite{WilsonEtal2017, BeutlerEtal2017B, CastorinaWhite2018, BeutlerCastorinaZhang2019}. This is given by
\be
    \tilde{P}(k) = \int \frac{dp}{2\pi^2} p^2 ~\mathcal{W}_{00}(k, p) P(p),
\ee
where $\mathcal{W}_{00}(k, p)$ is the mixing matrix 
\begin{equation}\label{W_00}
    \mathcal{W}_{00}(k, p) \equiv 4\pi \int dx ~ x^2 j_0(kx) j_0(px) Q(x),
\end{equation}
with $Q(x)$ being the 2PCF of the window. We refer to this approach as ``1DFFTLog approximation''. In this case the analysis is usually limited to those triangles where all sides are much larger than $1/R$, $R$ being the typical size of the window.

The two options described above apply to the tree-level bispectrum with the exception of the shot-noise contribution. The convolution of the latter is simply given by
\be
    \tilde{B}_{\mathrm{SN}}(\vec{p}_1, \vec{p}_2) = \frac{1}{\bar{n}}(1+\alpha_1)b_1^2\left[\tilde{P}_L(p_1) + \tilde{P}_L(p_2) + \tilde{P}_L(p_3)\right] + \frac{1}{\bar{n}^2}(1+\alpha_2),
\ee
where $\tilde{P}_L(p)$ is the window-convolved linear power spectrum, while it can be shown that the last constant term is not affected by the convolution.

As already mentioned, the ideal spherical geometry allows  the window 3PCF to be evaluated analytically. The window function in configuration space can be written as
\be \label{W_x_for_bisp}
    W(x) = \theta(R-x),
\ee
where $R$ is the radius of the sphere. The normalization is chosen such that the window 3PCF \eqref{Q_3pcf} is normalized to one for a uniform window function. The 3PCF multipoles $Q_\ell(x_{13}, x_{23})$ of the window function, assuming $x_{13} < x_{23}$, have the following form
\begin{figure}[h!]
    \includegraphics[width=0.99\textwidth]{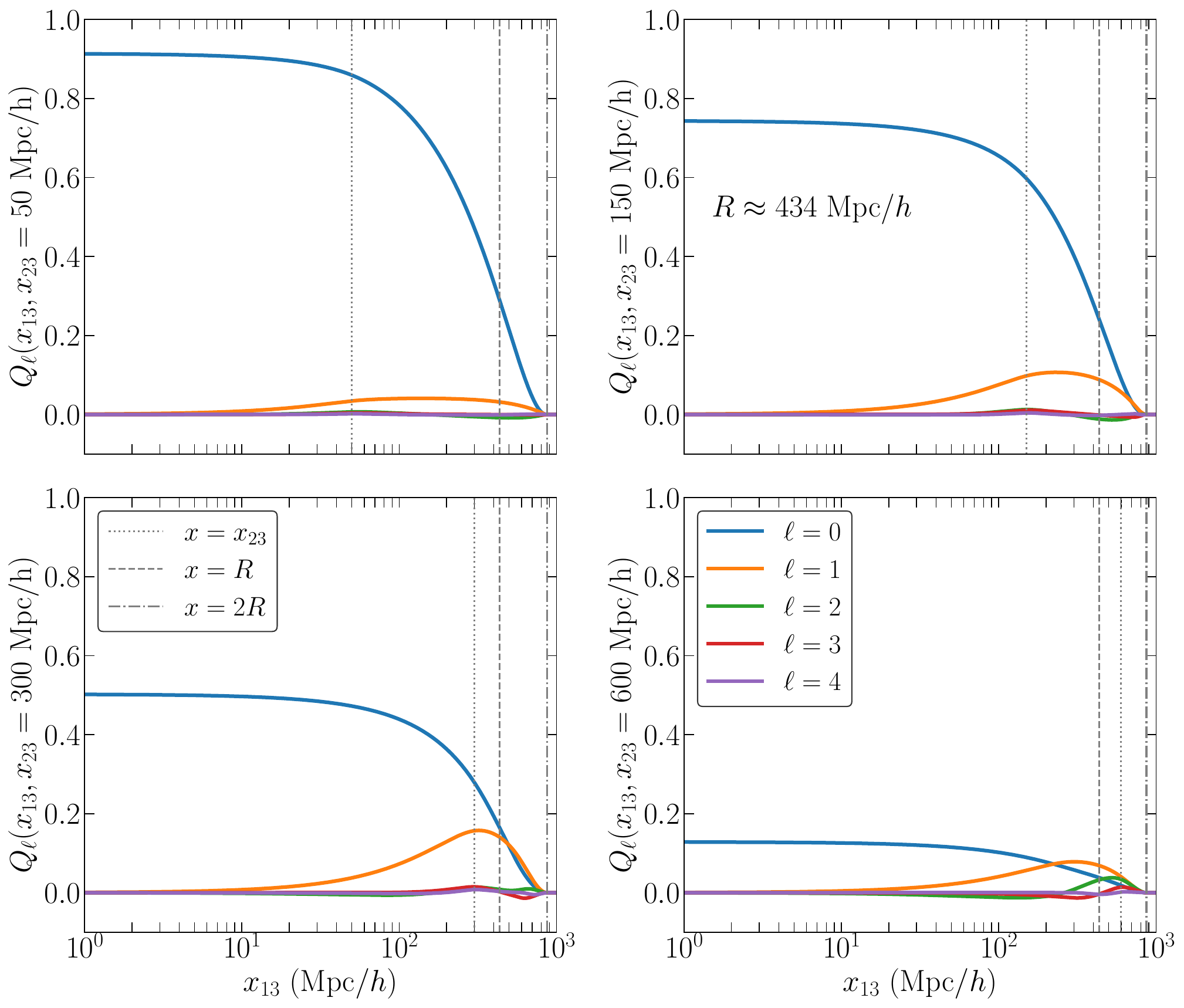}
    \centering
    \caption{Spherical window 3PCF multipoles $Q_\ell(x_{13},x_{23})$ plotted as a function of $x_{13}$ at different, constant values of $x_{23}$. It is evident that the relative contribution of higher-order $Q_{\ell}$ becomes quickly less significant as $\ell$ increases.}
    \label{fig:3pcf}
\end{figure}

\begin{equation}
\label{analytic_3pcf_expr}
Q_\ell \! = \!
    \begin{dcases}
    \left[
    \int_0^{R-x_{23}} \hspace{-2em} dx_{3} x_{3}^2  \int_{-1}^{1} \!\!\!\! d\mu_1 \!\int_{-1}^{1} \!\!\!\! d\mu_2
    +\!\!\int_{R -x_{23}}^{R-x_{13}} \hspace{-2em} dx_{3} x_{3}^2  \int_{-1}^{1} \!\!\!\! d\mu_1 \!\int_{-1}^{\mu_{2+}} \hspace{-1.5em} d\mu_2
    +\!\!\int_{R-x_{13}}^{R} \hspace{-2em} dx_{3} x_{3}^2  \!\int_{-1}^{\mu_{1+}} \!\!\!\! d\mu_1 \!\int_{-1}^{\mu_{2+}} \!\!\!\! d\mu_2
    \right]
    \int_{0}^{2\pi} \!\!\!\!\!\! d\phi \, G_\ell \\
    \hspace{26em} {\rm for}\quad x_{13} \le x_{23} \leq R \,, \\%%%%%%%%%%%%%%%%%%%%%%%%
    \left[
    \int_{x_{23} - R}^{R-x_{13}} \hspace{-2em} dx_{3} x_{3}^2  \!\int_{-1}^{1} \!\!\!\! d\mu_1 \!\int_{-1}^{\mu_{2+}} \!\!\!\! d \mu_2 +
     \int_{R -x_{13}}^{R} \hspace{-2em} dx_{3} x_{3}^2  \int_{-1}^{\mu_{1+}} \!\!\!\! d\mu_1 \int_{-1}^{\mu_{2+}} \!\!\!\! d \mu_2
    \right]\int_{0}^{2\pi} \!\!\!\! d\phi  \, G_\ell \\
    \hspace{22em} {\rm for} ~ x_{23} > R ~ \& ~ x_{13} + x_{23} \leq 2R\,,  
    \\ %%%%%%%%%%%%%%%%%%%%%%%%
    \int_{x_{23} - R}^{R} \hspace{-2em} dx_{3} x_{3}^2  \int_{-1}^{\mu_{1+}} \!\!\!\! d\mu_1 \int_{-1}^{\mu_{2+}} \!\!\!\! d\mu_2 \int_{0}^{2\pi} \!\!\!\! d\phi\, G_\ell
    \hspace{8.3em} {\rm for} ~ x_{23} > R ~ \& ~ x_{13} + x_{23} > 2R\,, \\ %%%%%%%%%%%%%%%%%%%%%%%%
    0 \hspace{25em} {\rm for}~ x_{23} > 2R \,,
    \end{dcases}       
\end{equation}
where
\be
G_\ell(\mu_1, \mu_2, \phi) \equiv \frac{2\ell+1}{2 V} \mathcal{L}_\ell\left[(1-\mu_1^2)^{1/2} (1-\mu_2^2)^{1/2} \cos \phi + \mu_1 \mu_2\right]\,,
\ee
while 
\be
\mu_{1+} \equiv (R^2 - x_{3}^2 - x_{13}^2)/(2 x_{3} x_{13})\,,~~
 {\rm and}~~ \mu_{2+} \equiv (R^2 - x_{3}^2 - x_{23}^2)/(2 x_{3} x_{23})\,.
 \ee

Slices of the window 3PCF multipoles $Q_{\ell}(x_{13}, x_{23})$, eq.~\eqref{analytic_3pcf_expr}, are shown in figure \ref{fig:3pcf}. By construction, from the choice of normalization in eq.~\eqref{W_x_for_bisp}, we see the monopole amplitude is always smaller than one. Note that the lower-order multipoles  dominate over the higher-order ones. In practice, as we will see later we only need to consider  multipoles of $Q_{\ell}(x_{13}, x_{23})$ up to $\ell = 2$ (see also figure~\ref{fig:var_others}).

The two-dimensional Hankel transform that is required to compute \eqref{W_llplpp} makes use of the public code \texttt{2DFFTLog}\footnote{\href{https://github.com/xfangcosmo/2DFFTLog}{github.com/xfangcosmo/2DFFTLog}} \cite{FangEiflerKrause2020}. We assume a bias $\nu(\ell) = 1.1$, a window filter of width = 0.25, without additional extrapolation nor padding. The details of the 2DFFTLog algorithm implementation are described in appendix~\ref{sys_test}, where we show how variations with respect to the reference set-up only induce $\lesssim 2\%$ differences on the output. For the 1DFFTLog approximation, we implement the one-dimensional Hankel transform in the mixing matrix of eq.~\eqref{W_00} using the code \texttt{mcfit}\footnote{\href{https://github.com/eelregit/mcfit}{github.com/eelregit/mcfit}} with a sampling of 10000 log-spaced points from $p_\mathrm{min} = 10^{-4}$ $h$/Mpc to $p_\mathrm{max} = 1$ $h$/Mpc that is further constantly padded using end-point of both sides, up to $2^{15}$ number of points.

%%%%%%%%%%%%%%%%%%%%%%%%%%%%%%%%%%%%%%%%%%%%%%%%%%%%%%%%%%%%%%%
\subsection{Binning effect}\label{binning_effect}

The measured $\hat{B}(k_1, k_2, k_3)$ is the average over all fundamental triangles $\{\qv_1, \qv_2, \qv_3\}$ that fall in the bins $q_i \in [k_i - \Delta k/2, k_i + \Delta k/2]$. Therefore our theoretical prediction for a given bin center $\{k_1, k_2, k_3\}$ should be binned in the same way.

In addition, a proper comparison with simulations  should account as well for the discrete nature of Fourier-space wavenumbers, taking values $\qv=\nv k_f$, an aspect that, for simplicity, we have not explicitly included in our definition of the estimator, eq.~\eqref{eq:BLest}.      

A binning operator acting on some function $F(\qv_1, \qv_2, \qv_3)$ should therefore be written as 
\be
\label{eq:binning_operator}
    \Big \langle F(\qv_1, \qv_2, \qv_3) \Big \rangle_\triangle \equiv \frac{1}{N_T} \sum_{\vec{q}_1 \in \vec{k}_1} \sum_{\vec{q}_2 \in \vec{k}_2} \sum_{\vec{q}_3 \in \vec{k}_3} F(\vec{q}_1, \vec{q}_2, \vec{q}_3) \delta_K(\vec{q}_{123})\,,
\ee
where $N_T(k_1, k_2, k_3)$ is the total number of ``fundamental'' triplets $\{\qv_1, \qv_2, \qv_3\}$ satisfying the condition $\qv_{123}=0$ enforced by the Kronecker symbol $\delta_K(\vec{q}_{123})$, that fall in the triangle bin $\{k_1, k_2, k_3\}$
\begin{equation}
    N_T(k_1, k_2, k_3) \equiv \sum_{\vec{q}_1 \in \vec{k}_1} \sum_{\vec{q}_2 \in \vec{k}_2} \sum_{\vec{q}_3 \in \vec{k}_3} \delta_K(\vec{q}_{123})\,.
\end{equation}
In the case of our 2DFFTLog approach, eq.~\eqref{2dfftlog-approach}, the effect of binning can be taken into account by applying such operator on the mixing matrix, eq.~\eqref{W_llplpp} as follows
\begin{align}
\label{2d_exact}
\mathcal{W}_{\ell \ell' \ell''}(k_1, k_2;p_1, p_2) \equiv & (4\pi)^2 (2\ell+1)\int d x_{13} x_{13}^2 \int d x_{23} x_{23}^2 ~ j_{\ell}(p_1 x_{13}) j_{\ell}(p_2 x_{23}) 
\nn \\ & \times 
\Big[ \Big \langle \mathcal{L}_{\ell''}(\hat{q}_1 \cdot \hat{q}_2)j_{\ell''}(q_1 x_{13}) j_{\ell''}(q_2 x_{23}) \Big\rangle_\triangle Q_{\ell'}(x_{13}, x_{23}) \Big]\,.  
\end{align}
The high dimensionality of the matrix makes this procedure very expensive to compute exactly according to eq.~\eqref{eq:binning_operator}. %Let us opt for a approximation for the exact binning. 
For this reason we consider two approximations explored already in \cite{OddoEtal2020, OddoEtal2021} (see also \cite{EggemeierEtal2021} and \cite{IvanovEtal2021A} for alternative approaches).

The first approximation amounts to simply evaluate our function \be
F(\qv_1, \qv_2, \qv_3) \equiv \mathcal{L}_{\ell}(\hat{q}_1 \cdot \hat{q}_2)j_{\ell}(q_1 x_{13}) j_{\ell}(q_2 x_{23})
\ee 
on the {\em sorted}, effective triplet $(\overline{k}_1, \overline{k}_2, \overline{k}_3)$ 
\begin{equation}\label{2d_eff}
    \Big \langle F(q_1, q_2, q_3) \Big \rangle_\triangle \simeq F(\overline{k}_1, \overline{k}_2, \overline{k}_3),
\end{equation}
where
\begin{align}\label{sorted_eff_tri}
    \nonumber \overline{k}_1 &\equiv \Big \langle \mathrm{max} \{q_1, q_2, q_3 \} \Big \rangle_\triangle, 
    \\ \overline{k}_2 &\equiv \Big \langle \mathrm{med} \{q_1, q_2, q_3 \} \Big \rangle_\triangle,
    \\ \nonumber \overline{k}_3 &\equiv \Big \langle \mathrm{min} \{q_1, q_2, q_3 \} \Big \rangle_\triangle,
\end{align}
such that $\mathrm{max} \{q_1, q_2, q_3\} \geq \mathrm{med} \{q_1, q_2, q_3\} \geq \mathrm{min} \{q_1, q_2, q_3\}$.
A better approximation can be obtained including higher-order corrections in a Taylor-expansion about the effective wavenumbers \cite{OddoEtal2021}. This corresponds to 
\begin{equation}\label{2d_expv2}
\Big \langle F(q_1, q_2, q_3) \Big \rangle_\triangle \simeq F(\overline{k}_1, \overline{k}_2, \overline{k}_3) + \frac{1}{2}\sum_{i, j} \frac{\partial^2}{\partial q_i \partial q_j} F(\overline{k}_1, \overline{k}_2, \overline{k}_3) \Big \langle  (q_i - \overline{k}_i)(q_j - \overline{k}_j) \Big \rangle_\triangle.
\end{equation}

On the other hand, the binning of the prediction in the 1DFFTLog approximation approach is closer to the procedure in \cite{OddoEtal2021} for the unconvolved bispectrum. For a generic tree-level expression such as $B(q_1, q_2, q_3)=K(\qv_1,\qv_2)P_L(q_1)P_L(q_1)+{\rm perm.}$, we evaluate the Taylor expansion around the $\overline{k}_i$ as 
\begin{equation}
\label{1d_exp}
\Big \langle \tilde{B}(q_1, q_2, q_3) \Big \rangle_\triangle \simeq \sum_{n+m\leq 2} \frac{1}{n! m!}\Big \langle K(\qv_1, \qv_2) (q_1 - \overline{k}_1)^n(q_2 - \overline{k}_2)^m \Big \rangle_\triangle \tilde{P}^{(n)}_L(\overline{k}_1) \tilde{P}^{(m)}_L(\overline{k}_2) + \mathrm{perm.}\,,
\end{equation}
where $\tilde{P}^{(n)}_L$ is the $n$-derivative of the convolved linear power spectrum. The $n=m=0$ term represent the leading contribution, corresponding to the first approximation of eq.~\eqref{2d_eff}.

%%%%%%%%%%%%%%%%%%%%%%%%%%%%%%%%%%%%%%%%%%%%%%%%%%%%%%%%%%%%%%%
\subsection{Likelihood analysis and covariance}

In the next section we test our convolution expression against measurements both from the Minerva simulations and from the \pin{} mocks. In both cases, the theoretical prediction is evaluated as the posterior-averaged convolved tree-level model by performing a likelihood analysis in terms of the five bias and shot-noise parameters, fixing the cosmological parameters. This consists in practice in averaging the model over the corresponding Markov chains.  

We consider a total log-likelihood function given by the sum of the log-likelihood of each individual realisation, so that
\be
\label{eq:likelihoodA}
\ln {\mathcal{L}}_{\rm tot} = \sum_{\alpha=1}^{N_R}\ln {\mathcal L}_{\rm \alpha}\,,
\ee
where $N_R$ is the total number of realisations of either mocks or simulations. For the single realisation we assume a Gaussian likelihood defined as
\be
\label{eq:likelihoodB}
    \ln \mathcal{L}_\alpha = -\frac{1}{2} \sum_{i, j} \Delta B_i C_{ij}^{-1} \Delta B_j, 
\ee
where $i$ is the triangle index, $\Delta B_i = \hat{B}_i^{(\alpha)} - \tilde{B}_i$ is the difference between the measurement and theoretical prediction, while $C_{ij}$ is the bispectrum covariance matrix estimated numerically from the $10000$ \pin{} mocks measurements. The large number of mocks provides a robust estimate of the covariance where residual statistical errors can be ignored \cite{OddoEtal2020}. For the maximum scale that we considered in our analysis $k_\mathrm{max} \approx 0.09\kMpc$, we have 143 triangles corresponding to a Hartlap factor \cite{HartlapSimonSchneider2007} of $\simeq 0.985$, implying an underestimation of variance of about 1$\%$.

\begin{figure}[t!]
    \includegraphics[width=0.99\textwidth]{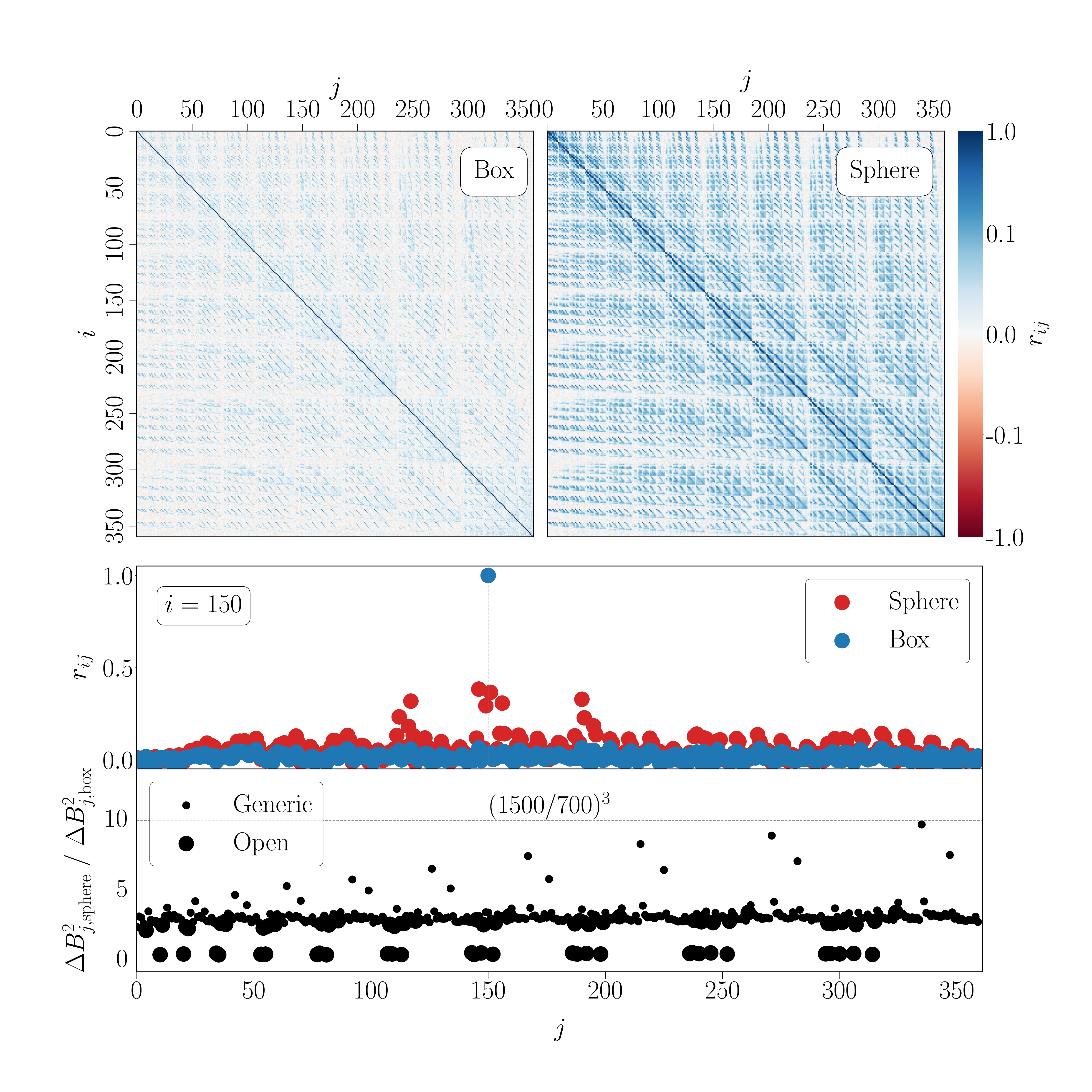}
    \centering
    \caption{The correlation matrix $r_{ij}$ for $\Delta k = 2 \, k_f$ bin size, estimated from 10000 \pin{} mocks. Mode-coupling from the window-bispectrum convolution induces a significant contribution from the non-diagonal term in the correlation matrix of the sphere data. The middle panel highlights the elements of $r_{ij}$ corresponding to the $i=150$ triangle. The bottom panel shows the ratio of the variance $\Delta B^2_i\equiv C_{ii}$ from the sphere and box catalogs. Leakage to the non-diagonal elements is the possible cause for this ratio to be less than the factor  expected from the volume ratio $\sim (1500/700)^3$ between the sphere and the box. 
    In particular, the effect is most severe for some of the open configurations, where the ratio of the variance is in the minimum. It is also interesting to note that the maximum of the ratio  happens to be in the squeezed configurations.}
    \label{fig:cov}
\end{figure}

The correlation matrix, defined as 
\be
r_{ij} \equiv \frac{C_{ij}}{\sqrt{C_{ii} C_{jj}}}\,,
\ee
estimated from the spherical window measurements is shown in figure~\ref{fig:cov} (top right panel), where it is compared to the analogous quantity corresponding to the box measurements (top left panel). The correlation among different modes induced by the window results clearly in larger off-diagonal elements. The middle panel of figure~\ref{fig:cov} shows the elements of $r_{ij}$ corresponding to the $i=150$ triangle, as a function of $j$, while the bottom panel shows the ratio of the variance $\Delta B^2_i\equiv C_{ii}$ from the sphere and box catalogs. Leakage to the non-diagonal elements likely causes this ratio to be less than the factor one could naively expect from the volume ratio $\sim (1500/700)^3$ between the sphere and the box.

We perform the analysis by means of MCMC (Monte Carlo Markov Chain) simulations using the Python implementation of affine-invariant ensemble sampler \texttt{emcee} \cite{ForemanMackeyEtal2013}. Initially, the linear power spectrum is obtained from the \texttt{CAMB} code \cite{LewisChallinorLasenby2000} assuming the same cosmology as the mocks. Each run of the chain, the bias parameter space is explored using 100 walkers. The chain is assumed to be converged after the number of steps is larger than 100 times integrated autocorrelation time $\tau$\cite{GoodmanWeare2010}. All the ingredients required to perform this analysis are compiled in the modified \texttt{PBJ} (P+B Joint analysis) code \cite{OddoEtal2020, OddoEtal2021}, which we extended to include the window-convolved bispectrum prediction. 
Note that in the case of fixed cosmology, the window-convolution process only needs to be done once, since the bias parameters can be factored-out.  Nevertheless, as we show in figure \ref{fig:time.pdf}  of appendix \ref{params_mem_time}, each evaluation of the convolved bispectrum is comparable with a typical call of the Boltzmann solver. More precisely, for bin size $\Delta k = 2 \, k_f$ with the total number of triangles $N_T = 360$, using as default parameters those described in appendix \ref{params_mem_time}, the evaluation of the tree-level model and the matrix multiplication take each $\sim 1$ second.

In all our fits, the free parameters are $\{b_1, b_2, b_{\mathcal{G}_2}, \alpha_1, \alpha_2\}$ with the priors given in table \ref{table:priors}.
\begin{table}[t!]
\centering
\begin{tabular}{cc}
Parameters                                                  & Priors (uniform) \\ \hline
\multicolumn{1}{c|}{$b_1$}                                  & {[}0.9, 3.5{]}  \\
\multicolumn{1}{c|}{$b_2$}                                  & {[}-4, 4{]}     \\
\multicolumn{1}{c|}{$b_{\mathcal{G}2}$} & {[}-5, 5{]}     \\
\multicolumn{1}{c|}{$\alpha_1$}                             & {[}-1, 1{]}     \\
\multicolumn{1}{c|}{$\alpha_2$}                             & {[}-1, 1{]}    
\end{tabular}
\caption{Prior assumed for the bias and shot-noise parameters in all likelihood evaluations.}
\label{table:priors}
\end{table}

\begin{comment}
\begin{figure}[h!]
    \includegraphics[width=12cm]{var.pdf}
    \centering
    \caption{Variance.}
    \label{fig:cov}
\end{figure}
\end{comment}

%%%%%%%%%%%%%%%%%%%%%%%%%%%%%%%%%%%%%%%%%%%%%%%%%%%%%%%%%%%%%%%
%%%%%%%%%%%%%%%%%%%%%%%%%%%%%%%%%%%%%%%%%%%%%%%%%%%%%%%%%%%%%%%
\section{Results}
\label{sec:results}

%%%%%%%%%%%%%%%%%%%%%%%%%%%%%%%%%%%%%%%%%%%%%%%%%%%%%%%%%%%%%%%
\subsection{Comparison with mocks measurements}

In this section we compare the prediction of the 2DFFTLog-based approach \eqref{2dfftlog-approach} and the approximation using 1DFFTLog \eqref{1dfftlog-approx} to the average of bispectrum measurements from the full set of \pin{} mocks. This extremely large set of measurements allows to keep the statistical noise below the systematic errors under investigation.  

Theoretical predictions are defined as the posterior-averaged bispectrum $\langle \tilde{B}(k_1,k_2,k_3) \rangle_{\rm post}$ from the likelihood analysis of the 10000 \pin{} box mocks including all triangles up to $k_\mathrm{max} = 0.09 \kMpc$, using the same data set to estimate the covariance. {\em Nota bene:} all theoretical predictions in this section, either for box or sphere measurements, are averaged over the same Markov chain corresponding to the likelihood analysis of the {\em box measurements}: this ensures that differences induced by systematic errors on the convolution are not reduced by fitting the free parameters. In other terms, we are comparing the different convolution methods adopting the model parameters determined, in both cases, from the analysis of the mocks bispectrum measured in the box according to the likelihood of eq's~\eqref{eq:likelihoodA} and \eqref{eq:likelihoodB}.  

The prediction for the unconvolved bispectrum is binned using the exact expression of eq.~\eqref{eq:binning_operator} (see also \cite{OddoEtal2020}), while the convolved prediction for both 2DFFTLog and 1DFFTLog approximation adopts the expansion method of eq.~\eqref{2d_expv2} and eq.~\eqref{1d_exp}, respectively. We have checked on a limited subset of configurations that even in the convolved bispectrum case, the difference between the expansion and the exact binning is below 0.5\%, more than an order of magnitude smaller than the effect of the window. 

We note in passing that the integral constraints correction in our setup is rather negligible. The correction term considered in \cite{PeacockNicholson1991, WilsonEtal2017} for the power spectrum, that would read here for the bispectrum as  $\tilde{B}(0,0,0) W(\vec{k}_1) W(\vec{k}_1) W(\vec{k}_3)$,  where $W(\vec{k})$ is the Fourier Transform of the properly normalized window function \eqref{W_x_for_bisp},  is less than $  4 \cdot 10^{-4}  \tilde{B}(0,0,0)$ for all the configurations considered. This is also due to the fact that the we excluded triangles with sides smaller than the fundamental frequency of the sphere $k_f^\mathrm{sphere}$.

\begin{figure}[t!]
    \includegraphics[width=0.95\textwidth]{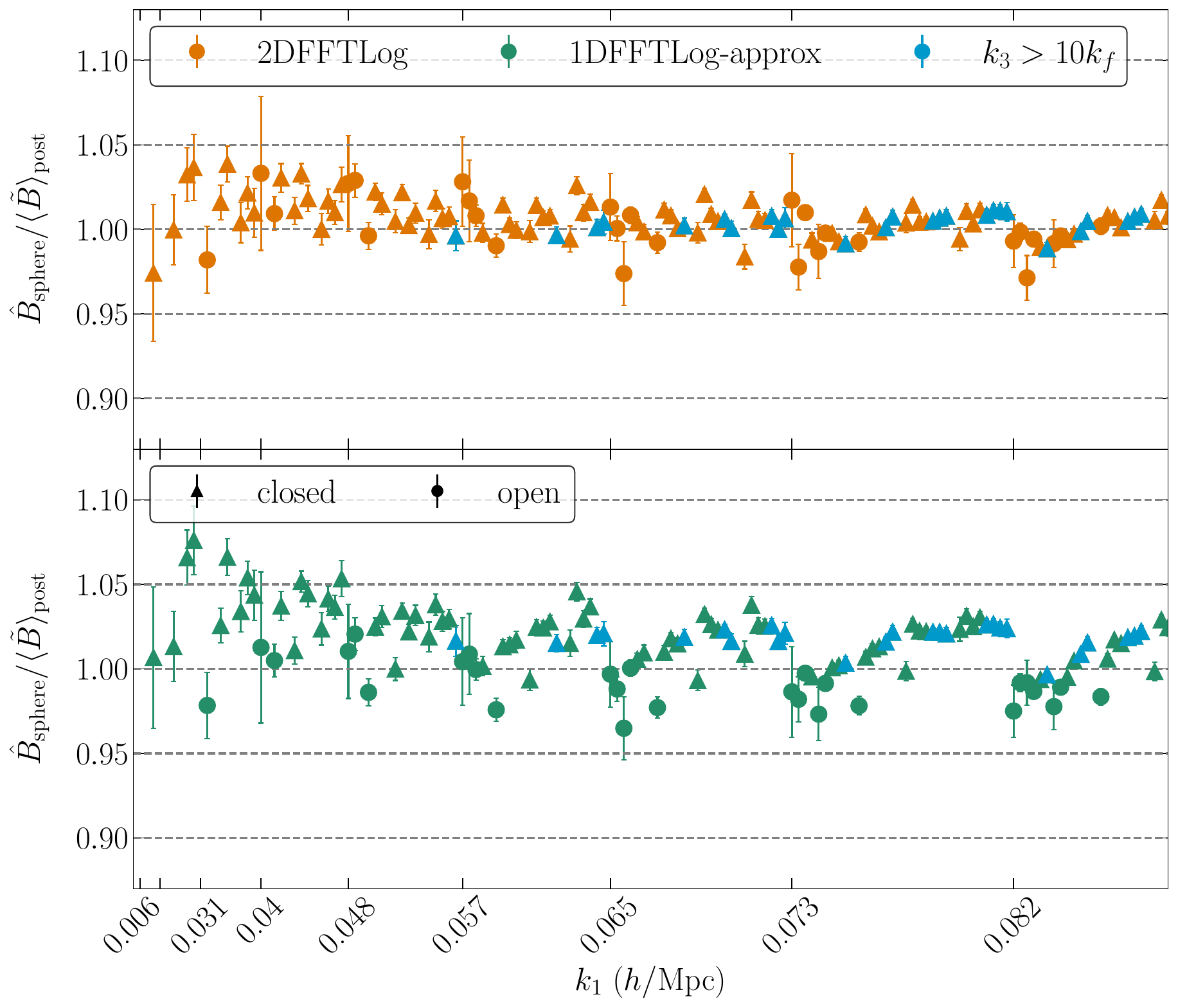}
    \centering
    \caption{Ratio of the average of the bispectrum measurements from the 10000 \pin{} mocks with spherical geometry for all configurations (up to $k_1 \leq 0.09$ $h$/Mpc) to the theoretical prediction of the 2DFFTLog method (top panel) and the 1DFFTlog approximation (bottom panel). Both  predictions are the (window-convolved) posterior-averaged bispectrum $\langle \tilde{B}(k_1,k_2,k_3) \rangle_{\rm post}$ from the likelihood analysis of the {\em box measurements} of the 10000 \pin{} mocks with $k_\mathrm{max} = 0.09 \kMpc$ using the same data set to estimate the covariance. The error bar is the error on the mean from the mocks. We mark in blue the configurations with the smallest wavenumber $k_3 > 10 k_f$, in general showing a better agreement between the prediction and mocks. }
    \label{fig:full_bisp_s2}
\end{figure}

Figure \ref{fig:full_bisp_s2} shows the ratio of the measurements to the convolved model for all configurations up to $k_{\rm max} = 0.09\kMpc$, with the top panel corresponding to the 2DFFTlog approach and the bottom panel to the 1DFFTlog approximation. The error bars correspond to the error on the mean of ratios of the 10000 \pin{} data and the prediction. The 2DFFTLog-based method matches the data fairly well with $\lesssim 2-3\%$ errors and a marginal, residual dependence on shape. The discrepancy becomes slightly worse on large scales but is still within $\lesssim 5\%$ error level. Here we observe a scatter among the very first points slightly larger then the estimated error bars. This could be due to the finite density of the random catalog adopted in the bispectrum estimator and to the fact that a single random has been used for all measurements. 

In the case of the 1DFFTLog approximation we notice differences with the measurements up to 6-7\% and a similar but larger shape-dependence. We also notice that at large scale, in addition to the scatter mentioned above, the 1DFFTLog method tends to underestimate the usual suppression of power. It should be noted, on the other hand, that the 1D approximation fares relatively well, despite the crude simplification.    

\begin{figure}[t!]
    \includegraphics[width=0.99\textwidth]{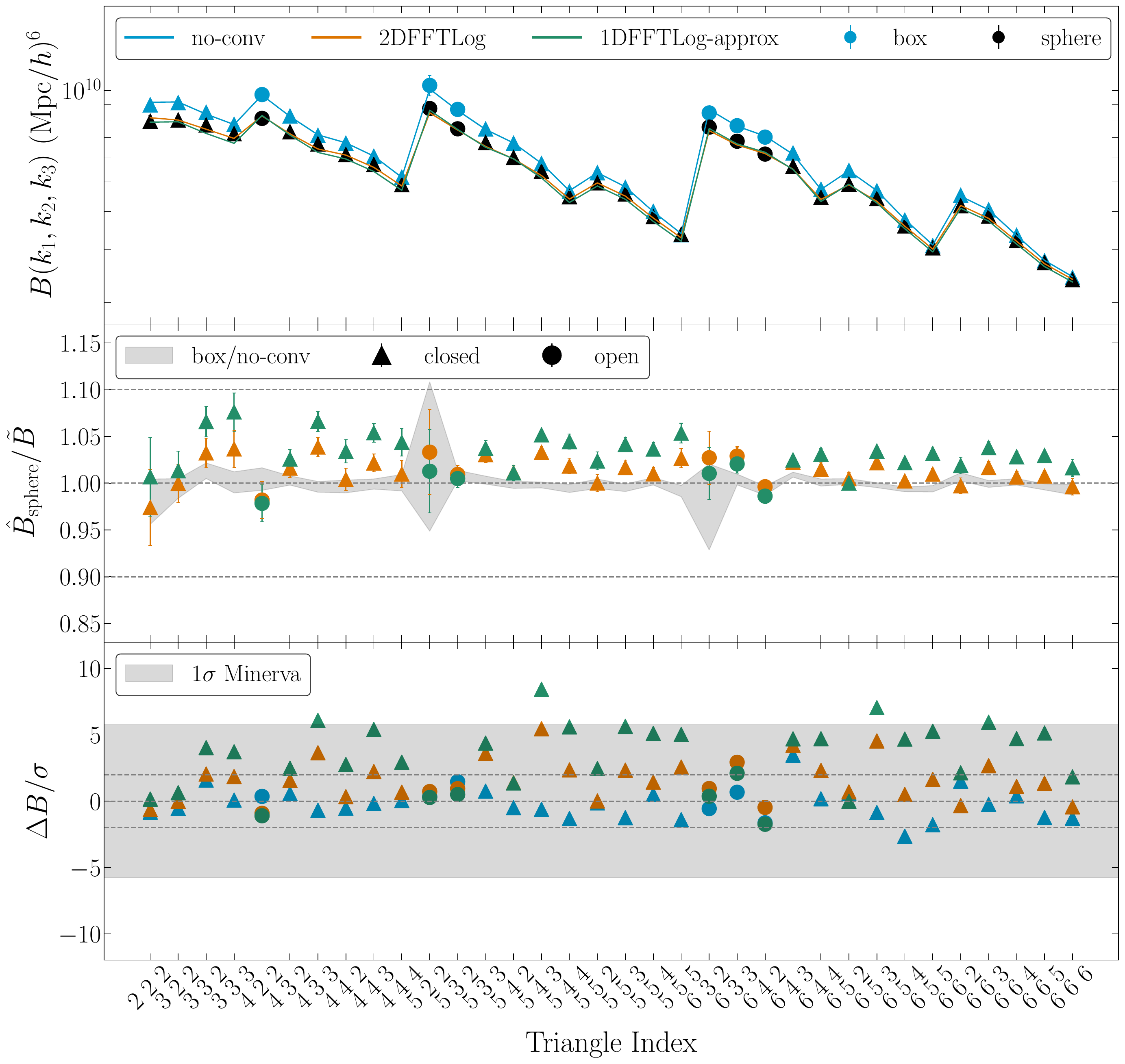}
    \centering
    \caption{Top panel: the measurements and the predictions, both in the box and the sphere, for the first 34 triangles of figure \ref{fig:full_bisp_s2}. All theoretical predictions are the (window-convolved) posterior-averaged bispectrum from the likelihood analysis of the 10000 \pin{} box mocks with $k_\mathrm{max} = 0.09 \kMpc$ using the same data set to estimate the covariance. Middle panel: the ratio between \pin{} mocks with spherical geometry measurements and the corresponding predictions, with the gray band corresponding to the 1-$\sigma$ uncertainty of the same ratio for the box/unconvolved case. Bottom panel: the relative error with respect to the variance of the 10000 \pin{} mocks. The 2DFFTLog method matches the data to a few times the level of uncertainty present in the unconvolved prediction (mostly within 2$\sigma$, with some outliers within 5$\sigma$). All theoretical prediction agree with the data within 1$\sigma$ corresponding to Minerva volume.}
    \label{fig:first_tris_s2}
\end{figure}

In figure~\ref{fig:full_bisp_s2} we also singled out, as blue points, those triangles where all sides are larger than $10 k_f \approx 0.04\kMpc $. This is approximately equivalent to the triangles selected for the analysis of \cite{GilMarinEtal2015, GilMarinEtal2015B}, where the 1DFFTLog method is applied. The motivation was to exclude the configurations where the effect of the window is not properly accounted for by the 1D approximation. Indeed, in the case of our spherical, ideal case, configurations excluding those pathological triangles have a lower systematic errors and smaller dependence on scale.

In figure \ref{fig:first_tris_s2} we zoom in on the first 34 triangles, roughly up to the largest side of the triangles $k_1 \approx 0.05\kMpc$. In the middle panel, we show the gray band corresponding to the 1-$\sigma$ uncertainty of the ratio between \pin{} mocks, box measurement and the unconvolved prediction. This shows that part of the scatter and to some extent, the shape-dependent systematic error is also present in the box comparison and it is therefore not due to the convolution. We see as well that the 2DFFTLog method matches the data fairly well, at the level of uncertainty present in the box case. The 1DFFTLog approximation is $\sim$ (1-2)$\sigma$ worse than the 2DFFTLog and is consistently underpredicting the data. The third panel shows that the 2DFFTLog prediction is mostly within 2$\sigma$ (comparable with the unconvolved prediction) with some outliers still within 5$\sigma$. The panel also shows as a gray band the 1-$\sigma$ uncertainty corresponding to the measurements from the set of Minerva simulations, that we will consider in the next section.

In the end, in our test, we see that window effect on the bispectrum is a very tiny effect mainly on large scales which can be only probed by using a very large number of realisations. We  notice that the 1DFFTlog approximation fares rather well, with a small overall offset that does not present a marked dependence on shape, at least in our spherical, ideal case in real-space. Still, we will study how such offset, present with the same size and sign on all configurations, could affect the analysis of data set even of much smaller cumulative volumes than the extreme case considered here.

%%%%%%%%%%%%%%%%%%%%%%%%%%%%%%%%%%%%%%%%%%%%%%%%%%%%%%%%%%%%%%%
\subsection{Recovering bias parameters from the N-body measurements}

We consider now {\em independent fits} to the N-body simulations of the theoretical model with each of the convolution methods in order to assess their effect on the recovered bias parameters. Again we fit individually all distinct realisations according to the likelihood function described in eq.s~\eqref{eq:likelihoodA} and \eqref{eq:likelihoodB}.

The sphere measurements from the 298 Minerva simulations correspond to a more realistic test since the total volume, of about $100\cGpc$, is roughly 1/6 of the blinded challenge volume presented in \cite{NishimichiEtAl2020} and approximately 6 times the volume of the largest redshift bin of Euclid survey (the $(z_\mathrm{min}, z_\mathrm{max}) = (1.5,1.8)$ redshift bin which total volume is $16.22 \cGpc$) considered in the forecasts of \cite{EuclidIST:F2019}. As it can be seen from the bottom panel of figure \ref{fig:first_tris_s2}, both the unconvolved and window-convolved prediction agree with the data within 1$\sigma$ of the Minerva measurements. Still, a small systematic effect distributed over a large number of data-points can affect the determination of cosmological parameters. 

\begin{figure}[t!]
    \includegraphics[width=0.99\textwidth]{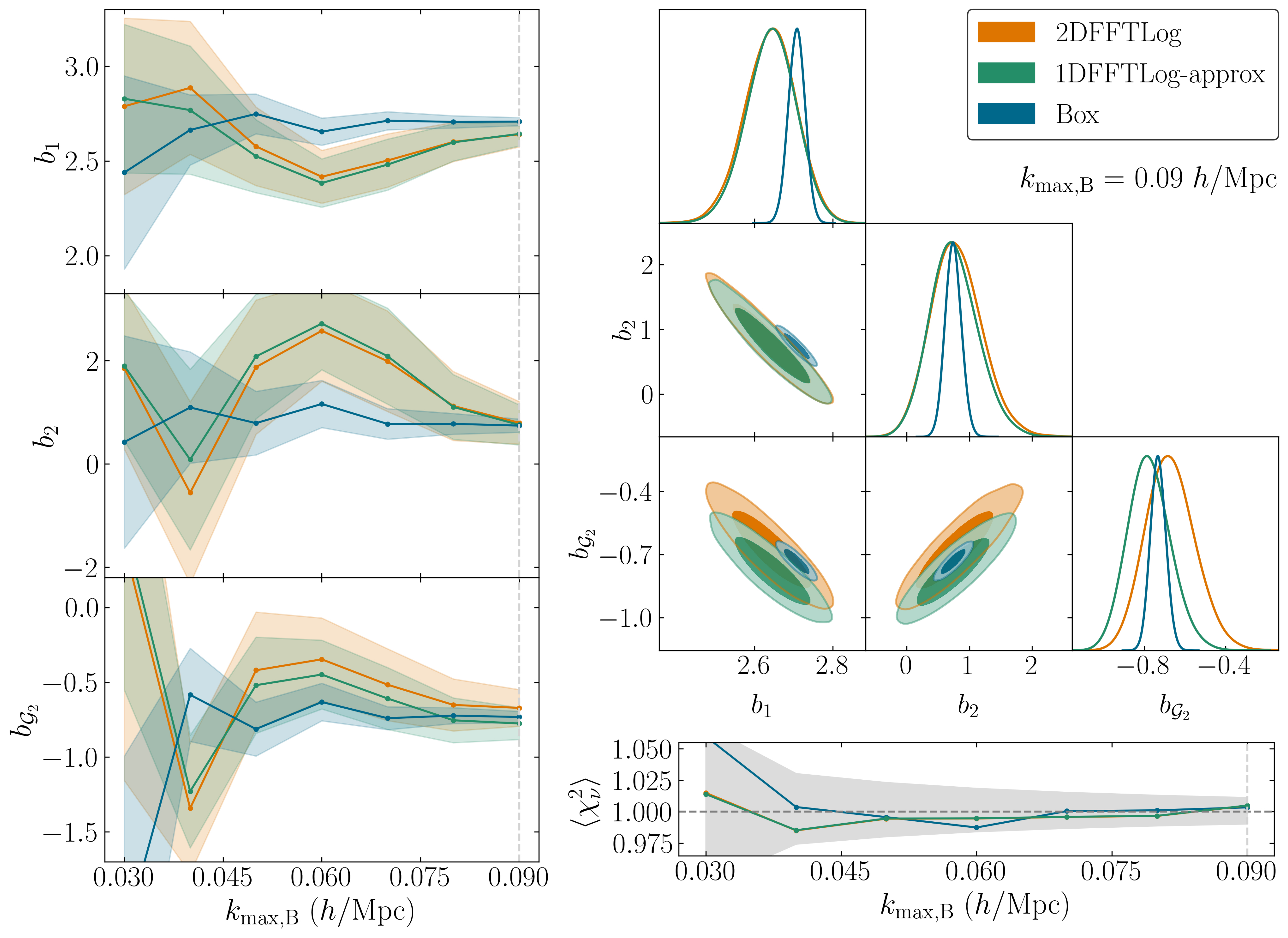}
    \centering
    \caption{{\em Left panels}: marginalised 1-$\sigma$ constraints on the bias parameters as a function of $k_{\rm max}$ from the box measurements analysis (blue) and from the sphere measurements assuming the 2DFFTlog (orange) and 1DFFTlog approximation (green) predictions. {\em Top-right panel}: marginalised 2-parameters contours at $k_{\rm max}=0.09\kMpc$. {\em Bottom-left panel}: posterior-averaged, reduced  chi-square $\langle \chi_\nu^2 \rangle$ as a function of $k_{\rm max}$, with the gray band marking the 95\% C.L. interval.}
    \label{fig:contour_s2}
\end{figure}

The posteriors for the bias parameters are shown in figure \ref{fig:contour_s2}. The figure provides the 1-$\sigma$ constraints as a function of the largest side of the triangle on the left panels while the top-right panel shows the marginalised 2-parameters contours at $k_{\rm max}=0.09\kMpc$. Finally, the bottom-left panel shows the posterior-averaged, reduced chi-square $\langle \chi_\nu^2 \rangle$ as a function of $k_{\rm max}$, compared with the 95\% C.L. interval shown as a gray band. The larger constraints from the Minerva sphere analysis with respect to the Minerva box analysis are simply due to the difference in volume, which a ratio of roughly a factor of 8. All constraints are marginalised over the shot-noise parameters that we are not showing for clarity.   

We can see that bias parameters recovered from Minerva box and sphere analysis (for both methods) agree within 1$\sigma$. This level of agreement is expected since as we have shown in the previous section, sample variance dominates over window effects on large scales. Yet, in the contour plots one can notice a purely systematic effect almost at the 2-$\sigma$ level between the two convolution methods that could become more relevant at smaller scales, where the model should be extended to include loop corrections. 

We derived all the results presented in this section assuming a different, smaller bin size ($\Delta k = k_f$) for the wavenumber in all bispectrum measurements. Since we do not observe any relevant difference in our conclusions we do not include them in this presentation.

%%%%%%%%%%%%%%%%%%%%%%%%%%%%%%%%%%%%%%%%%%%%%%%%%%%%%%%%%%%%%%%
%%%%%%%%%%%%%%%%%%%%%%%%%%%%%%%%%%%%%%%%%%%%%%%%%%%%%%%%%%%%%%%
\section{Conclusions}
\label{conclusion}

In this work we laid out an expression for the exact convolution with a survey window function for the redshift-space bispectrum multipoles as defined in \cite{Scoccimarro2015}. The window convolution we propose is based on a two-dimensional Hankel transform implemented by means of the \texttt{2DFFTlog} code of \cite{FangEiflerKrause2020}, an extension of the 1D FFTlog algorithm by \cite{Hamilton2000}. Our implementation allows for an efficient computation, via FFTs, of the mixing matrix relating theoretical predictions to the corresponding observable by means of a matrix  multiplication. The mixing matrix has to be computed only once and it is provided as input to the likelihood analysis\footnote{In our case the evaluation of the mixing matrix takes a few hours (see table \ref{table:mem_time_eff_exp})}. The multiplication of the theory bispectrum with the mixing matrix, in the case of a total number of triangles $N_T = 360$, takes $\sim 1$ second which is comparable to the running time of a typical call to a Boltzmann solver. On the other hand our approach provides an improvement over the 1D approximation adopted in \cite{GilMarinEtal2015, GilMarinEtal2015B, GilMarinEtal2017, DAmicoEtal2020, DAmicoEtal2022A} and should provide a benchmark for alternative bispectrum estimators such as those of \cite{SugiyamaEtal2019} and \cite{Philcox2021}. 

We test our expression in the ideal setting of a spherical window function where the 3-point correlation function of the footprint can be computed analytically. We compare our prediction for the convolved bispectrum monopole in real space to the average of measurements from halo mock catalogs. The average is taken from an extremely large number of realizations, in order to keep the statistical noise below the systematic errors under consideration. We notice that, at least in this example, the effect of the window is quite small. Nevertheless, our 2DFFTlog approach reduces significantly the errors characterising the 1DFFTLog approximation, with residuals at the few-percent level and weak dependence on shape. Always within the spherical window model, we then explored the effect of the two methods on the determination of the bias (and shot-noise) parameters from the halo catalogs of the Minerva simulations for a total volume of about 100$\cGpc$. We find differences in the determination of the linear and nonlocal bias parameters at the 2-$\sigma$ level, suggesting the need of further investigations in a more realistic context.    

The results of this preliminary comparison might quantitatively change in more realistic setups.  For instance, the effect of the window will be more evident for higher multipoles in redshift-space and for models with a specific signal in squeezed configuration such as local primordial non-Gaussianity or relativistic effects \cite{UmehEtal2017, KaragiannisEtal2018, ClarksonEtal2019, DeWeerdEtal2020, MaartensEtal2020, Barreira2020, MoradinezhadDizgahEtal2021, EnriquezHidalgoValenzuela2021A, MaartensEtal2021, Barreira2022}. Also, while the overall window effect should decrease at small scales ($k\gtrsim 0.1\kMpc$), the 1D approximation is expected to fail when loop corrections become important: in this respect, again, squeezed configuration might require particular attention.  Finally, the 1D approximation does not take into account the coupling of different bispectrum multipoles induced by window function effects. 

We will explore in future work more realistic scenarios. Here, we limit ourselves to observe that the implementation of the multiplication by the window matrix introduced here does not significantly changes the run-time for a  typical Markov chain and could therefore provide a safe and robust approach to the bispectrum convolution for the bispectrum multipole estimator of \cite{Scoccimarro2015}.           

\acknowledgments

We are always grateful to Claudio Dalla Vecchia and Ariel Sanchez for running and making available the Minerva simulations, performed on the Hydra and Euclid clusters at the Max Planck Computing and Data Facility (MPCDF) in Garching.
We would like to thank Davit Alkhanishvili, Alexander Eggemeier, Chiara Moretti for valuable discussions and Andrea Oddo for the help with \texttt{PBJ} code. We also would like to thank the anonymous referee for the very careful reading of our manuscript and the many valuable suggestions that surely improved our presentation. The \pin{} mocks were run on the GALILEO cluster at CINECA, thanks to an agreement with the University of Trieste.
K.P., E.S.\ and P.M.\ are partially supported by the INFN INDARK PD51 grant and acknowledge support from PRIN MIUR 2015 Cosmology and Fundamental Physics: illuminating the Dark Universe with Euclid. M.B.\ acknowledges support from the Netherlands Organization for Scientific Research (NWO), which is funded by the Dutch Ministry of Education, Culture and Science (OCW) under VENI grant 016.Veni.192.210.

%%%%%%%%%%%%%%%%%%%%%%%%%%%%%%%%%%%%%%%%%%%
\appendix

%%%%%%%%%%%%%%%%%%%%%%%%%%%%%%%%%%%%%%%%%%%
%%%%%%%%%%%%%%%%%%%%%%%%%%%%%%%%%%%%%%%%%%%
\section{Convergence tests and binning systematics}
\label{sys_test}

The evaluation of the mixing matrix underlying all the results presented in the main text adopts a set of default values for several parameters. We explore in section~\ref{ssec:convergence} how our results depend on their choice, while in section~\ref{ssec:binning} we quantify the systematic error related to the binning scheme. 

%%%%%%%%%%%%%%%%%%%%%%%%%%%%%%%%%%%%%%%%%%%
\subsection{2DFFTLog parameters}
\label{ssec:convergence}

The first relevant parameter in the 2DFFTLog evaluation of eq.~\eqref{W_llplpp} is the number of sampling points $N_p$ for $p_1$ and $p_2$, i.e. the number of points where the theory bispectrum is evaluated (eq.~\eqref{2dfftlog-approach}).
The prediction for the unconvolved bispectrum is evaluated for all (closed) triangles one can form based on the $N_p$ wavenumber values, so that one of the dimension of the mixing matrix scales like $N_p^3$. We assume as default value $N_p=512$. Other two parameters are the maximum values for the multipoles $\ell$ and $\ell'$, for which we take as default $\ell_\mathrm{max}=30$ and $\ell'_\mathrm{max}=2$. We have then the range of integration over $p$, with $\{p_{\rm min},p_{\rm max}\}=\{10^{-5},0.5\}\kMpc$ as reference.   

\begin{figure}[t!]
    \includegraphics[width=0.85\textwidth]{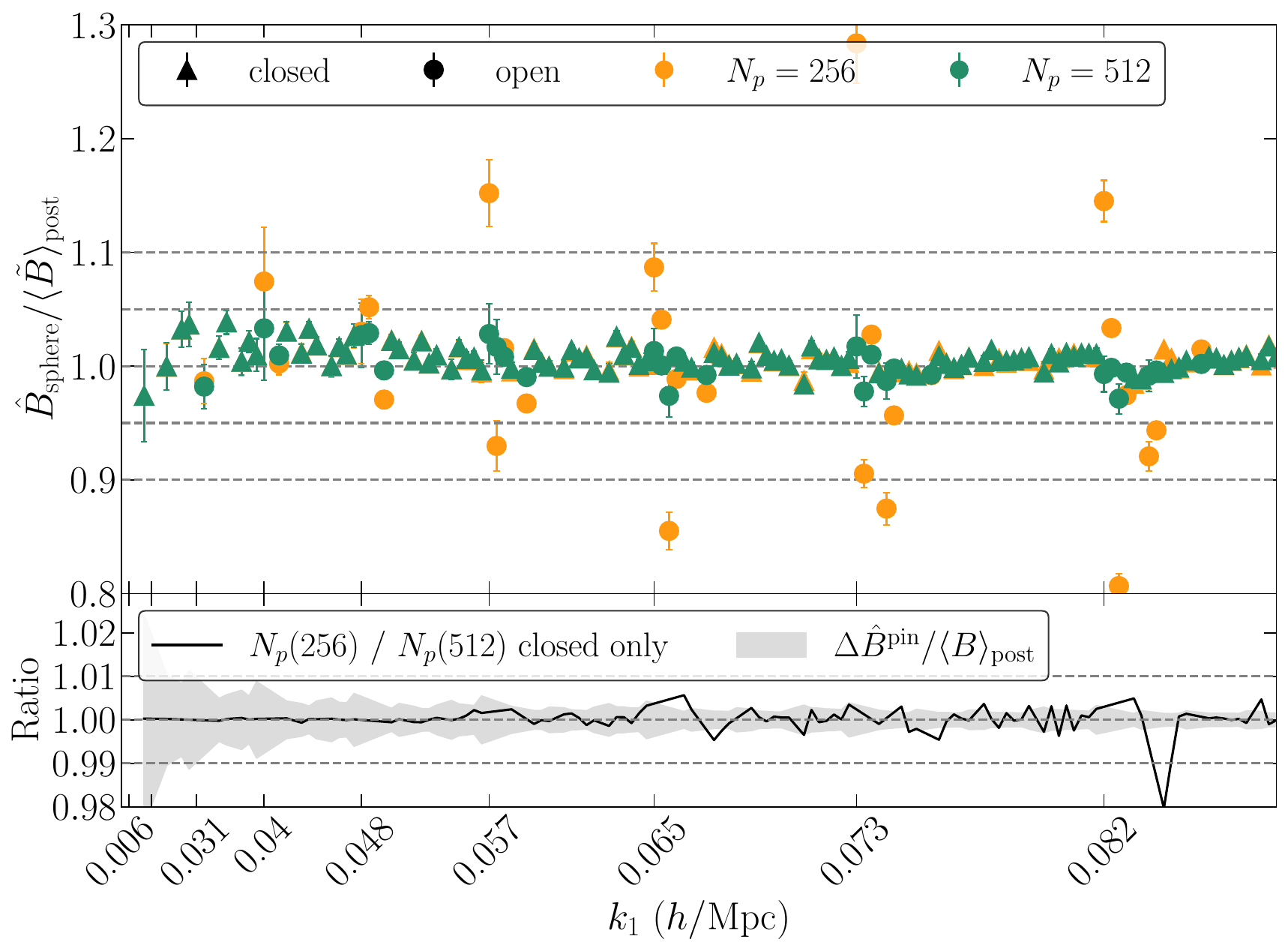}
    \centering
    \caption{Top panel: the effect of varying the number of sampling points $N_p$ in the comparison of measurements (10000 \pin{} mocks with spherical geometry) to the theoretical prediction (window-convolved posterior-averaged from the likelihood analysis of the 10000 \pin{} box mocks) evaluated for $N_p=512$ (the default value, green) and $N_p=256$ (yellow). ``Open'' triangles are shown with circle markers while regular, ``closed'' triangles are shown with triangles. Bottom panel: ratio of the two predictions compared to the 1-$\sigma$ error on the mean of the \pin{} box measurements $\Delta \hat{B}^\mathrm{pin}$ divided by the posterior-averaged of the unconvolved theory $\langle B\rangle_\mathrm{post}$.}
    \label{fig:var_nump}
\end{figure}

The effect of varying the number of sampling points $N_p$ (fixing the other parameters to their default values) is shown in figure \ref{fig:var_nump}. Interestingly enough, reducing $N_p$ to $256$ affects significantly the open configurations, while it only gives a $\leq 2\%$ difference on regular, closed triangles. Reducing $N_p$ has the advantage of reducing the mixing matrix generation time significantly (see table \ref{table:mem_time_eff_exp}) while also reducing the time required by the matrix multiplication (see figure 
\ref{fig:time.pdf}). A practical approach could be a hybrid method where the 1DFFTLog approximation handles the open configurations (where based on figure \ref{fig:full_bisp_s2}, only induce systematics within 5$\%$) while the 2DFFTLog is used for the rest.

\begin{figure}[t!]
    \includegraphics[width=0.9\textwidth]{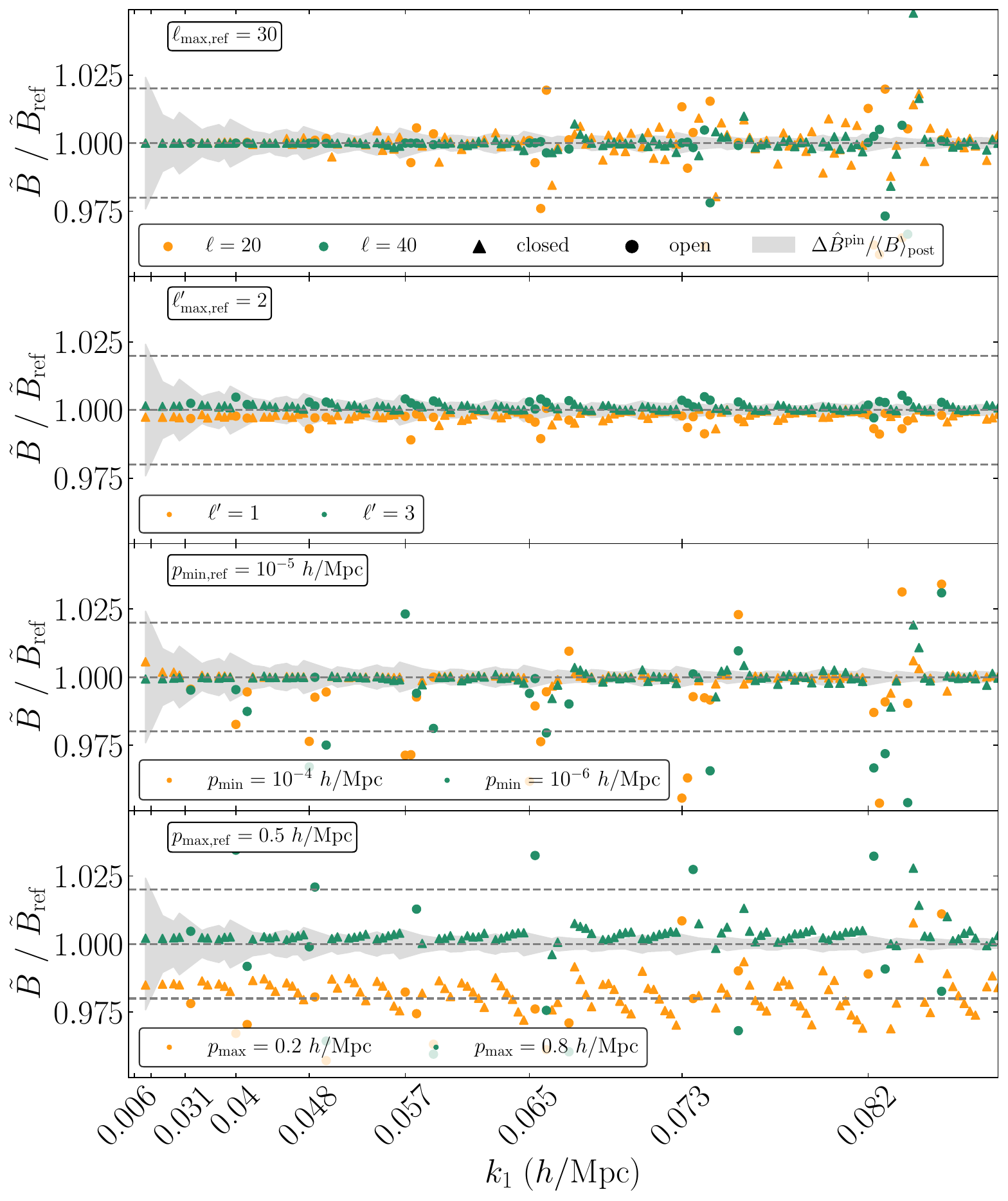}
    \centering
    \caption{Ratio of the prediction for the window-convolved  bispectrum (posterior-averaged from the likelihood analysis of the 10000 \pin{} box mocks) computed with different choices for the parameters $\ell_{\rm max}$, $\ell'_{\rm max}$, $p_{\rm min}$ and $p_{\rm max}$ w.r.t. the reference prediction (which again is the posterior-averaged). The gray band is the 1-$\sigma$ error on the mean of the \pin{} box measurements $\Delta \hat{B}^\mathrm{pin}$ divided by the posterior-averaged of the unconvolved theory $\langle B\rangle_\mathrm{post}$.}
    \label{fig:var_others}
\end{figure}

Figure \ref{fig:var_others} shows the effects of varying the other four parameters $\ell_{\rm max}$, $\ell'_{\rm max}$, $p_{\rm min}$ and $p_{\rm max}$, in terms of ratios to the default evaluation of the prediction. Varying $p_\mathrm{min}$ and $p_\mathrm{max}$ is done in such a way that the spacing $\Delta \ln p$ is fixed. The variation of all these parameters only induce $\lesssim 2\%$ difference with respect to the default case except for $p_\mathrm{max} = 0.2\kMpc$ which apparently is too low for our data vector extending to $k_\mathrm{max} \approx 0.09$ $h$/Mpc.

%%%%%%%%%%%%%%%%%%%%%%%%%%%%%%%%%%%%%%%%%%%
\subsection{Binning effects}
\label{ssec:binning}

\begin{figure}[t!]
    \includegraphics[width=0.9
\textwidth]{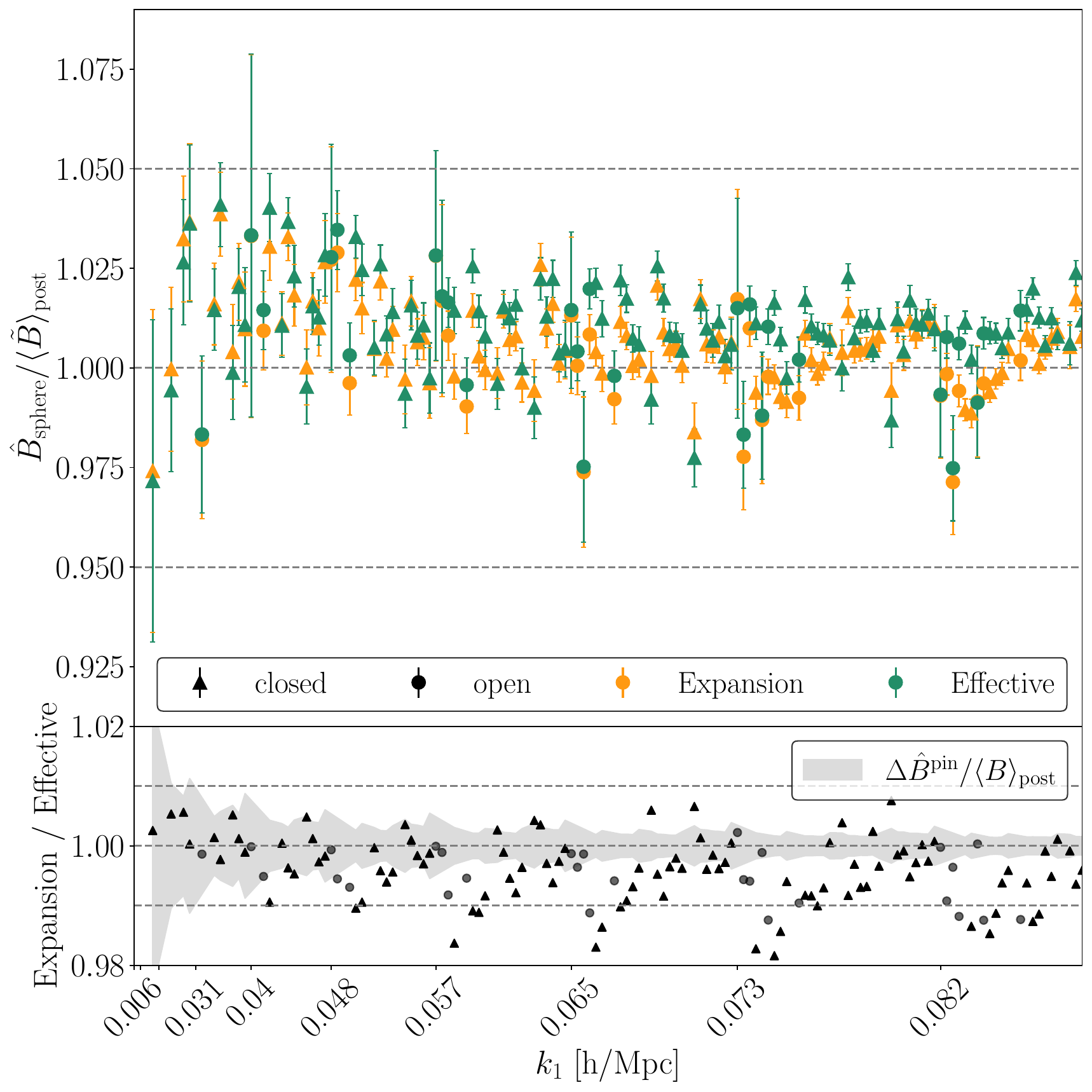}
    \centering
    \caption{Top panel: Comparison between the simpler effective binning of eq.~\eqref{2d_eff} with expansion binning of eq.~\eqref{2d_expv2} for $\Delta k=2 \, k_f$ bin size. Predictions are the window-convolved posterior-averaged bispectrum from the likelihood analysis of the 10000 \pin{} box mocks, while the measurements come from the 10000 \pin{} mocks with spherical geometry. Bottom panel: ratio of the two binning methods compared to the 1-$\sigma$ error on the mean of the \pin{} box measurements $\Delta \hat{B}^\mathrm{pin}$ divided by the posterior-averaged of the unconvolved theory $\langle B\rangle_\mathrm{post}$. The difference is only at $\lesssim 2\%$ level.}
    \label{fig:var_bin}
\end{figure}

In figure~\ref{fig:var_bin}, we compare the simpler effective binning of eq.~\eqref{2d_eff} with the Taylor expansion defined in eq.~\eqref{2d_expv2}, which is used for all our results in the main text. The effect of binning for the window-convolved bispectrum is at the $\lesssim 2\%$ level, which is smaller than effect of binning on the typical unconvolved bispectrum \cite{OddoEtal2020}. 
The advantage of the effective binning is again that it speeds up the computation of the mixing matrix (see table \ref{table:mem_time_eff_exp}). As an additional check, we also computed the exact binning procedure of eq.~\eqref{2d_exact} for a limited set of representative triangular configurations. We found that the difference with the Taylor expansion approach is below $0.5\%$ for all triangles considered. 

%%%%%%%%%%%%%%%%%%%%%%%%%%%%%%%%%%%%%%%%%%%%%%%%%%%%%%%%%%
%%%%%%%%%%%%%%%%%%%%%%%%%%%%%%%%%%%%%%%%%%%%%%%%%%%%%%%%%%
\section{Memory and time requirement} 
\label{params_mem_time}

\begin{figure}[t!]
    \includegraphics[width=0.75\textwidth]{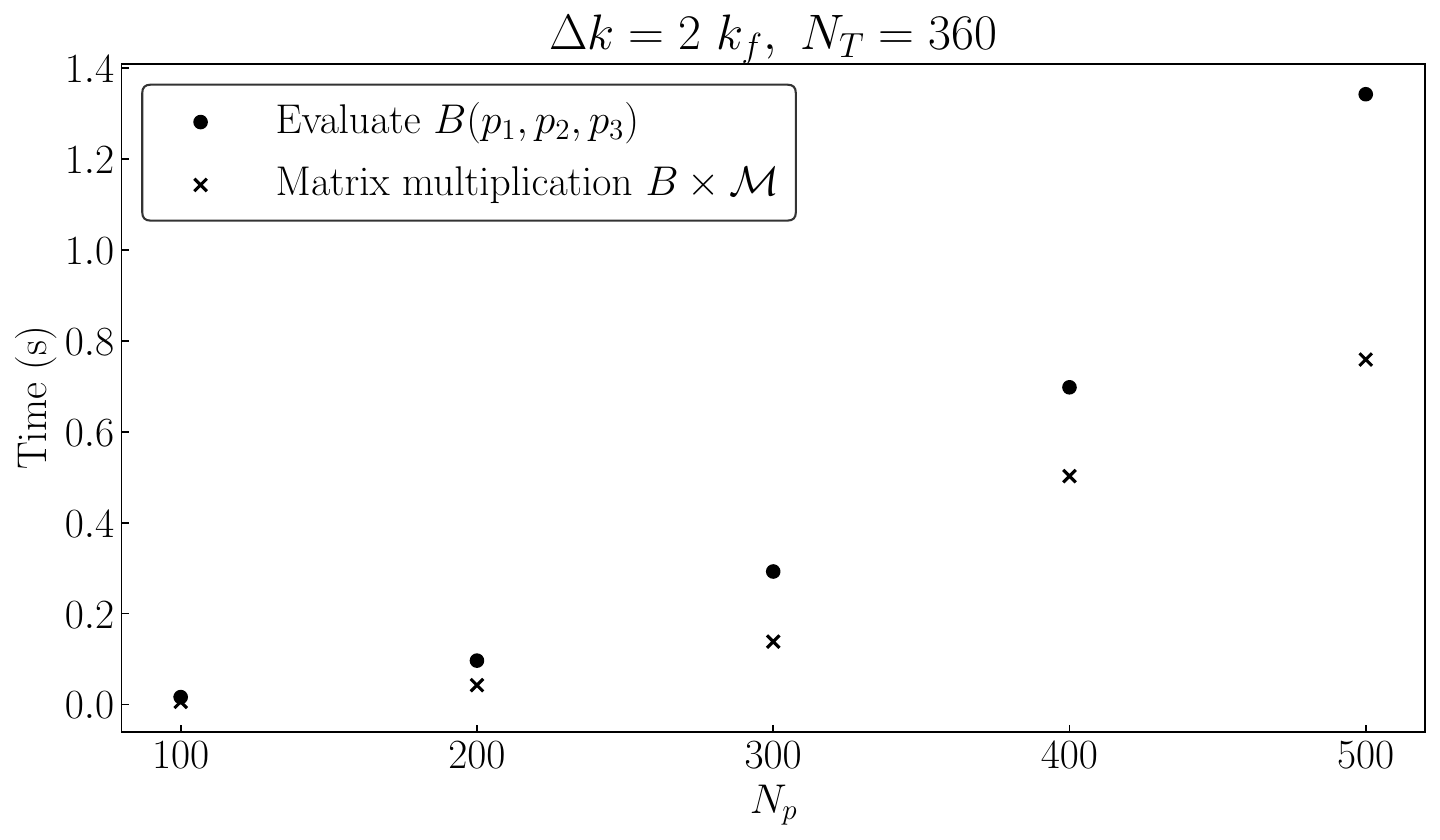}
    \centering
    \caption{Time required for the matrix multiplication compared to the evaluation of the unconvolved bispectrum $B(p_1,p_2,p_3)$ as a function of the number of sampled points $N_p$. In all cases the number of measured triangles is kept fixed at $N_T=360$. The code is run in the similar infrastructure as in table~\ref{table:mem_time_eff_exp}. Our code doesn't make use of parallel computing.}
    \label{fig:time.pdf}
\end{figure}

In this appendix we describe the performance of the Python implementation of the 2DFFTLog method for the bispectrum convolution in our test of a real-space, spherical window function. In table \ref{table:mem_time_eff_exp} we compare the time and memory requirement to generate the mixing matrix, for the two binning methods of eq.s~ \eqref{2d_eff} and \eqref{2d_expv2} and for different values of the number of sampling points $N_p$\footnote{The code is run on a single node on the \textsc{Ulysses} cluster v2 at SISSA. The nodes we use are equipped with $10$ Intel(R) Xeon(R) E5-2680 v2 $@$ 2.80GHz CPUs with $40$ GB of RAM in total.}. In general, the expansion method takes a factor 2 or 3 more time to generate the mixing matrix, while requiring approximately the same amount of memory.
The time required as a function of $N_p$ is also shown in figure \ref{fig:time.pdf}. The initialization of each chain in a MCMC run, which consists of generating the triangles $\{p_1, p_2, p_3\}$ in log-space and loading the mixing matrix $\mathcal{M}$, takes few minutes. At each sampling of the parameter space that requires a call for the linear power spectrum, the unconvolved bispectrum has to be evaluated on the triangles  $\{p_1, p_2, p_3\}$ and then multiplied with the mixing matrix. Both processes separately take $\sim$ 1 second for $N_p\sim 500$. The time taken by the matrix multiplication also depends linearly, as one can expect, on the number of redshift-space multiples considered and on the number of triangles $N_T$, which can be reduced by choosing a larger bin size $\Delta k$.

\begin{table}[t!]
\centering
\begin{tabular}{cccc}
\multicolumn{4}{c}{$\Delta k=2 \, k_f$,~ $N_T = 360$}                                                                               \\ \hline
\multicolumn{1}{c|}{Binning method}             & \multicolumn{1}{c|}{$N_p$} & \multicolumn{1}{c|}{Memory (GB)} & Time (hrs:mins) \\ \hline
\multicolumn{1}{c|}{\multirow{2}{*}{Effective}} & \multicolumn{1}{c|}{512}   & \multicolumn{1}{c|}{29.85}       & 02:01           \\
\multicolumn{1}{c|}{}                           & \multicolumn{1}{c|}{256}   & \multicolumn{1}{c|}{6.55}        & 00:15           \\ \hline
\multicolumn{1}{c|}{\multirow{2}{*}{Expansion}} & \multicolumn{1}{c|}{512}   & \multicolumn{1}{c|}{30.08}       & 04:02           \\
\multicolumn{1}{c|}{}                           & \multicolumn{1}{c|}{256}   & \multicolumn{1}{c|}{7.02}        & 00:45          
\end{tabular}
\caption{Comparison of memory and time required to generate the mixing matrix for different binning methods and different values of the number of sampling points $N_p$. The code is run on a single node in Ulysses cluster v2 at SISSA which is equipped with 10 CPU Intel(R) Xeon(R) E5-2680 v2 @ 2.80GHz and with 40 GB of RAM in total. Our code however doesn't make use of parallel computing.}
\label{table:mem_time_eff_exp}
\end{table}

%%%%%%%%%%%%%%%%%%%%%%%%%%%%%%%%%%%%%%%%%%%%%%%%%%%%%%%%%%%%%%%%
%%%%%%%%%%%%%%%%%%%%%%%%%%%%%%%%%%%%%%%%%%%%%%%%%%%%%%%%%%%%%%%%
%\section{On the window 3PCF multipoles $Q^L_{L'M' \ell \ell' \ell_1 \ell_2}(x_{13}, x_{23})$}
\section{On the measurement of the window 3PCF multipoles}
\label{app:window3PCF}

We are not addressing in this work the problem of the optimal estimator for the window 3PCF multipoles $Q^L_{L'M' \ell \ell' \ell_1 \ell_2}(x_{13}, x_{23})$ defined in eq.~\eqref{3PCF_multipoles_redshifts}, and of the systematic errors related to their measurement. We simply notice that in addition to direct counting (which even for the case of anisotropic 3PCF multipoles has been shown can be reduced to pair counting complexity \cite{Slepian:2017lpm}), one could consider an FFT-based estimator, since 
\begin{align}
Q^{L}_{L' M'  \ell \ell' \ell_1 \ell_2 }(x_{13}, x_{23}) =\, & (-1)^{M'}\sum_{\substack{M, m_1, m_2\\ m, m'}}
\sum_{\substack{\tilde{\ell}_1, \tilde{\ell}_2 \\ \tilde{m}_1, \tilde{m}_2}}
4\pi i^{\ell' - \ell + \ell_2 - \ell_1}
~\mathcal{G}_{L \ell_1 \ell_2 }^{M m_1 m_2} \mathcal{G}_{L' \ell \ell'}^{M' m m'}
\mathcal{G}_{\tilde{\ell}_1 \ell_1 \ell'}^{\tilde{m}_1 m_1 m'}
\mathcal{G}_{\tilde{\ell}_2 \ell_2 \ell}^{\tilde{m}_2 m_2 m}
\nn \\ & \times 
\int_V ~d^3 x_3 ~W_{\tilde{\ell}_1 \tilde{m}_1}(\vec{x}_3; x_{13}) W_{\tilde{\ell}_2 \tilde{m}_2}(\vec{x}_3; x_{23}) W_{LM}(\vec{x}_3)\,,
\end{align}
can be obtained in terms of the Fourier Transform
\be
    W_{\ell m}(\vec{x}_3;x_{ij}) \equiv   
    \, i^{\ell} \int \frac{d^3 q}{(2\pi)^3} e^{i \vec{q} \cdot \vec{x}_3} j_{\ell}(q x_{ij}) Y_{\ell m}(\hat{q}) W(\vec{q})\,, 
\ee
and 
\be
   W_{LM}(\vec{x}_3) \equiv  \,  W(\vec{x_3}) Y^*_{LM}(\vec{x}_3)\,.
\ee

The window 3PCF 
\begin{equation}
    Q(\vec{x}_{13}, \vec{x}_{23}, \hat{x}_3) \equiv \frac1{V}\int d x_3 x_3^2 ~ W(\vec{x}_{13} + \vec{x}_3) W(\vec{x}_{23} + \vec{x}_3) W(\vec{x}_3),
\end{equation}
can also be expanded in TripoSH basis with zero total angular momentum,
as follows \cite{SugiyamaEtal2019}
\begin{align}
Q(\vec{x}_{13}, \vec{x}_{23}, \hat{x}_3) = \sum_{\ell_1, \ell_2, L} Q_{\ell_1 \ell_2 L}(x_{13}, x_{23}) \times &\Bigg[\sqrt{\frac{4\pi}{(2\ell_1+1)(2\ell_2+1)(2L+1)}} \begin{pmatrix}
\ell_1 & \ell_2 & L\\
0 & 0 & 0
\end{pmatrix}^{-2} \nn \\& \, \, \sum_{m_1, m_2, M}  \mathcal{G}_{\ell_1 \ell_2 L}^{m_1 m_2 M}  Y_{\ell_1 m_1}(\hat{x}_{13}) Y_{\ell_2 m_2}(\hat{x}_{23}) Y_{LM}(\hat{x}_{3})\Bigg],
\end{align}
where then by construction
\begin{align}
    Q_{\ell_1 \ell_2 L}(x_{13}), x_{23}) = \sum_{m_1, m_2, M} 
    &\begin{pmatrix}
    \ell_1 & \ell_2 & L \\
    0 & 0 & 0
    \end{pmatrix}
    \begin{pmatrix}
    \ell_1 & \ell_2 & L \\
    m_1 & m_2 & M
    \end{pmatrix}
    \\&\int d^2 \hat{x}_3 \int d^2 \hat{x}_{13} \int d^2 \hat{x}_{23} ~Q(\vec{x}_{13}, \vec{x}_{23}, \hat{x}_3) Y_{\ell_1 m_1}^*(\hat{x}_{13}) Y_{\ell_2 m_2}^*(\hat{x}_{23}) Y_{L M}^*(\hat{x}_3).
\end{align}
The window 3PCF multipoles, eq.~\eqref{3PCF_multipoles_redshifts}, are then related to the $Q_{\ell_1 \ell_2 L}(x_{13}, x_{23})$, via the following linear combination 
\begin{equation}
Q^L_{L' M' \ell \ell' \ell_1 \ell_2}(x_{13}, x_{23}) = \sum_{\tilde{\ell}_1,\tilde{\ell_2}} Q_{\tilde{\ell}_1 \tilde{\ell}_2 L}(x_{13}, x_{23}) \,\mathbb{M}_{\tilde{\ell}_1 \tilde{\ell}_2 L; L' M' \ell \ell' \ell_1 \ell_2},
\end{equation}
where
\begin{align}
\mathbb{M}_{\tilde{\ell}_1 \tilde{\ell}_2 L; L' M' \ell \ell' \ell_1 \ell_2} \equiv & \sum_{\substack{M, m_1, m_2\\ m, m', \tilde{m}_1, \tilde{m}_2}} (-1)^{M+\tilde{m}_1+\tilde{m}_2}  \,i^{\ell' - \ell + \ell_2 - \ell_1} \sqrt{\frac{1}{4\pi (2\tilde{\ell}_1+1)(2\tilde{\ell}_2+1)(2L+1)}} 
\nn \\ & \times
\begin{pmatrix}
\tilde{\ell}_1 & \tilde{\ell}_2 & L\\
0 & 0 & 0
\end{pmatrix}^{-2}
 \, \mathcal{G}_{L \ell_1 \ell_2 }^{M m_1 m_2} \, \mathcal{G}_{L \tilde{\ell}_1 \tilde{\ell}_2 }^{M -\tilde{m}_1 -\tilde{m}_2} \, \mathcal{G}_{L' \ell \ell'}^{M' m m'} \, \mathcal{G}_{\ell_1 \ell' \tilde{\ell}_1 }^{m_1 m' \tilde{m}_1} \, \mathcal{G}_{\ell_2 \ell \tilde{\ell}_2 }^{m_2 m \tilde{m}_2}\,.
\end{align}
Written this way, the corresponding window bispectrum multipoles $B_{W, \ell_1 \ell_2 L}(k_1, k_2)$, which are related to $Q_{\ell_1 \ell_2 L}(x_{13}, x_{23})$ through a double Hankel transform \cite{SugiyamaEtal2019}, can also be used as an input measurement (see also \cite{Umeh:2020zhp} for similar application of double Hankel transform to relate the bispectrum multipoles to the 3PCF multipoles).

%%%%%%%%%%%%%%%%%%%%%%%%%%%%%%%%%%%%%%%%%%%%%%%%%%%%%%%%%%
%%%%%%%%%%%%%%%%%%%%%%%%%%%%%%%%%%%%%%%%%%%%%%%%%%%%%%%%%%
\section{Derivation of the real-space window convolution expressions in section \ref{sec:real-space-window}}
\label{sec:real-space-window-derivation}

Taking $L=0, M'=0, L'=0$
\begin{align}
Q^{0}_{0 0  \ell \ell' \ell_1 \ell_2 }(x_{13}, x_{23}) \equiv &   \frac{1}{\sqrt{4\pi}}\sum_{\tilde{\ell}_1, \tilde{\ell}_2} \sum_{\substack{m_1, m_2\\ m, m', \tilde{m}_1, \tilde{m}_2}} 4\pi i^{\ell' - \ell + \ell_2 - \ell_1} \, \mathcal{G}_{0 \ell_1 \ell_2 }^{0 m_1 m_2} \, \mathcal{G}_{0 \ell \ell'}^{0 m m'} \, \mathcal{G}_{\ell_1 \ell' \tilde{\ell}_1 }^{m_1 m' \tilde{m}_1} \, \mathcal{G}_{\ell_2 \ell \tilde{\ell}_2 }^{m_2 m \tilde{m}_2} \,
\nonumber \\ & \times
\int_V d^3 x_3 \int \frac{d^2 \hat{x}_{13}}{4\pi} \int \frac{d^2 \hat{x}_{23}}{4\pi} Y_{\tilde{\ell}_1 \tilde{m}_1}(\hat{x}_{13}) Y_{\tilde{\ell}_2 \tilde{m}_2}(\hat{x}_{23})
\nonumber \\ & \times
W(\vec{x}_3 + \vec{x}_{13}) W(\vec{x}_3 + \vec{x}_{23}) W(\vec{x}_3),
\end{align}
as $Y_{00}(\hat{x}_3) = 1/\sqrt{4\pi}$ in our convention.
From $\mathcal{G}_{0 \ell_1 \ell_2}^{0 m_1 m_2}= \tfrac{(-1)^{m_1}}{\sqrt{4\pi}}\delta_{\ell_1 \ell_2} \delta_{m_1, -m_2}$, we have
\begin{align}
Q^{0}_{0 0  \ell \ell' \ell_1 \ell_2 }(x_{13}, x_{23}) \equiv &   \frac{1}{\sqrt{4\pi}}\sum_{\tilde{\ell}_1, \tilde{\ell}_2} \sum_{\substack{m_1, m'\\ \tilde{m}_1, \tilde{m}_2}} \, (-1)^{m_1+ m'} \delta_{\ell_1 \ell_2} \delta_{\ell \ell'}\, \mathcal{G}_{\ell_1 \ell' \tilde{\ell}_1 }^{m_1 m' \tilde{m}_1} \, \mathcal{G}_{\ell_1 \ell' \tilde{\ell}_2 }^{-m_1, -m', \tilde{m}_2} \,
\nonumber \\ & \times
\int_V d^3 x_3 \int \frac{d^2 \hat{x}_{13}}{4\pi} \int \frac{d^2 \hat{x}_{23}}{4\pi} Y_{\tilde{\ell}_1 \tilde{m}_1}(\hat{x}_{13}) Y_{\tilde{\ell}_2 \tilde{m}_2}(\hat{x}_{23})
\nonumber \\ & \times
W(\vec{x}_3 + \vec{x}_{13}) W(\vec{x}_3 + \vec{x}_{23}) W(\vec{x}_3).
\end{align}
From the fact that $m_1 + m' +\tilde{m}_1 =0$ and $\ell_1 +\ell'+\tilde{\ell}_2$ is even, we have
\begin{align}
    \sum_{m_1, m'} (-1)^{m_1 + m'} \mathcal{G}_{\ell_1 \ell' \tilde{\ell}_1}^{m_1 m' \tilde{m}_1} \mathcal{G}_{\ell_1 \ell' \tilde{\ell}_2}^{-m_1, -m', \tilde{m}_2} &= \sum_{m_1, m'} (-1)^{\tilde{m}_1} \frac{(2\ell_1+1)(2\ell'+1)}{4\pi} \sqrt{(2\tilde{\ell}_1 + 1)(2\tilde{\ell}_2+1)}
    \nonumber \\ &
    \begin{pmatrix}
    \ell_1 & \ell' & \tilde{\ell}_1 \\
    0 & 0 & 0
    \end{pmatrix}
    \begin{pmatrix}
    \ell_1 & \ell' & \tilde{\ell}_1 \\
    m_1 & m' & \tilde{m}_1
    \end{pmatrix}
    \begin{pmatrix}
    \ell_1 & \ell' & \tilde{\ell}_2 \\
    0 & 0 & 0
    \end{pmatrix}
    \begin{pmatrix}
    \ell_1 & \ell' & \tilde{\ell}_2 \\
    m_1 & m' & -\tilde{m}_2
    \end{pmatrix}
    \nonumber \\ &
    = (-1)^{\tilde{m}_1} \frac{(2 \ell_1 +1)(2\ell' +1)}{4\pi} 
    \begin{pmatrix}
    \ell_1 & \ell' & \tilde{\ell}_1 \\
    0 & 0 & 0
    \end{pmatrix}^2
    \delta_{\tilde{\ell}_1 \tilde{\ell}_2} \delta_{\tilde{m}_1, -\tilde{m}_2},
\end{align}
where orthogonality of 3$j$-symbols
\begin{equation}
    \sum_{m_1, m'} (2\tilde{\ell}_1+1) \begin{pmatrix}
    \ell_1 & \ell' & \tilde{\ell}_1 \\
    m_1 & m' & \tilde{m}_1
    \end{pmatrix}
    \begin{pmatrix}
    \ell_1 & \ell' & \tilde{\ell}_2 \\
    m_1 & m' & -\tilde{m}_2
    \end{pmatrix}=\delta_{\tilde{\ell}_1 \tilde{\ell}_2} \delta_{\tilde{m}_1, -\tilde{m}_2},
\end{equation}
has been used, so that
\begin{align}
Q^{0}_{0 0  \ell \ell' \ell_1 \ell_2 }(x_{13}, x_{23}) \equiv & \frac{1}{\sqrt{4\pi}}   \sum_{\tilde{\ell}_1, \tilde{\ell}_2} \sum_{\tilde{m}_1, \tilde{m}_2} \, \frac{1}{4\pi} \delta_{\ell_1 \ell_2} \delta_{\ell \ell'} \delta_{\tilde{\ell}_1,\tilde{\ell}_2} \delta_{\tilde{m}_1, -\tilde{m}_2}\, 
(2\ell_1+1)(2\ell'+1) 
\begin{pmatrix}
\ell_1 & \ell' & \tilde{\ell}_1 \\
0 & 0 & 0
\end{pmatrix}^2 \,
\nonumber \\ & \times
\int_V d^3 x_3 \int \frac{d^2 \hat{x}_{13}}{4\pi} \int \frac{d^2 \hat{x}_{23}}{4\pi} Y_{\tilde{\ell}_1 \tilde{m}_1}(\hat{x}_{13}) Y^*_{\tilde{\ell}_1 \tilde{m}_1}(\hat{x}_{23})
\nonumber \\ & \times
W(\vec{x}_3 + \vec{x}_{13}) W(\vec{x}_3 + \vec{x}_{23}) W(\vec{x}_3)
\nonumber \\& 
= \frac{1}{\sqrt{4\pi}}\sum_{\tilde{\ell}_1} \frac{1}{(4\pi)^2}\, \delta_{\ell_1 \ell_2} \delta_{\ell \ell'} (2 \ell_1 + 1) (2\ell'+1) (2\tilde{\ell}_1+1)
\begin{pmatrix}
\ell_1 & \ell' & \tilde{\ell}_1 \\
0 & 0 & 0
\end{pmatrix}^2
\nonumber \\& \times
\int_V d^3 x_3 \int \frac{d^2 \hat{x}_{13}}{4\pi} \int \frac{d^2 \hat{x}_{23}}{4\pi} \mathcal{L}_{\tilde{\ell}_1}(\hat{x}_{13} \cdot \hat{x}_{23}) 
W(\vec{x}_3 + \vec{x}_{13}) W(\vec{x}_3 + \vec{x}_{23}) W(\vec{x}_3).
\end{align}

The matrix $\mathcal{Q}^{L}_{L' M' \ell}(q_1, q_2, q_3; p_1, p_2)$ then becomes
\begin{align}
\mathcal{Q}^{0}_{0 0 \ell}(q_1, q_2, q_3; p_1, p_2) &\simeq \sum_{\ell_1,\ell_2, \ell'} 16 \pi^2 \frac{I_{\ell_2 \ell_2 0}(q_1, q_2, q_3)}{I_{000}(q_1, q_2, q_3)} \mathcal{W}^0_{0 0 \ell \ell' \ell_1 \ell_2}(q_1, q_2; p_1, p_2)
\nonumber \\&= 
\frac{1}{\sqrt{4 \pi}} \sum_{\ell_1,\tilde{\ell}_1} 16 \pi^2 (-1)^{\ell_1} \mathcal{L}_{\ell_1}(\hat{q}_1 \cdot \hat{q}_2) \times \frac{1}{(4\pi)^2} (2\ell_1+1)(2\ell+1)
\begin{pmatrix}
\ell_1 & \ell & \tilde{\ell}_1 \\
0 & 0 & 0
\end{pmatrix}^2
\nonumber \\& \times
(4\pi)^2 \int dx_{13} x_{13}^2 \int dx_{23} x_{23}^2 \, j_{\ell}(p_1 x_{13}) j_{\ell}(p_2 x_{23}) j_{\ell_1}(q_1 x_{13}) j_{\ell_1}(q_2 x_{23})
\nonumber \\& \times
\left[ (2\tilde{\ell}_1+1) \int_V d^3 x_3 \int \frac{d^2 \hat{x}_{13}}{4\pi} \int \frac{d^2 \hat{x}_{23}}{4\pi}  
W(\vec{x}_3 + \vec{x}_{13}) W(\vec{x}_3 + \vec{x}_{23}) W(\vec{x}_3) \mathcal{L}_{\tilde{\ell}_1}(\hat{x}_{13} \cdot \hat{x}_{23}) \right]
\nonumber \\& =
\frac{1}{\sqrt{4 \pi}} \sum_{\tilde{\ell}_1, \ell_1} (2\ell_1+1)(-1)^{\ell_1} 
\begin{pmatrix}
\ell & \tilde{\ell}_1 & \ell_1 \\
0 & 0 & 0
\end{pmatrix}^2
\nonumber \\& \times 
(4\pi)^2 (2\ell+1) \int dx_{13} x_{13}^2 \int dx_{23} x_{23}^2 \, j_{\ell}(p_1 x_{13}) j_{\ell}(p_2 x_{23})
\nonumber \\& \times 
\left[\mathcal{L}_{\ell_1} (\hat{q}_1 \cdot \hat{q}_2) j_{\ell_1}(q_1 x_{13}) j_{\ell_1}(q_2 x_{23})  Q_{\tilde{\ell}_1}(x_{13}, x_{23})\right],
\end{align}

where 
\begin{equation}
Q_{\tilde{\ell}_1}(x_{13}, x_{23}) \equiv (2\tilde{\ell}_1 +1) \int_V d^3 x_3 \int \frac{d^2 \hat{x}_{13}}{4 \pi} \int \frac{d^2 \hat{x}_{23}}{4 \pi} W(\vec{x}_3 + \vec{x}_{13}) W(\vec{x}_3 + \vec{x}_{23}) W(\vec{x}_3) \mathcal{L}_{\tilde{\ell}_1}(\hat{x}_{13} \cdot \hat{x}_{23}).
\end{equation}
Identifying $(\ell, \tilde{\ell}_1, \ell_1) = (\ell, \ell', \ell'')$ and defining 
\begin{align}
\mathcal{Q}_{\ell} (q_1, q_2, q_3;p_1, p_2) &\equiv \sqrt{4\pi} \mathcal{Q}_{0 0 \ell}^0(q_1, q_2, q_3; p_1, p_2),
\\ 
\mathcal{W}_{\ell \ell' \ell''}(q_1, q_2;p_1, p_2) & \equiv  (4\pi)^2 (2\ell+1)\int d x_{13} x_{13}^2 \int d x_{23} x_{23}^2 ~ j_{\ell}(p_1 x_{13}) j_{\ell}(p_2 x_{23}) 
\nonumber \\ & \times 
\Big[  \mathcal{L}_{\ell''}(\hat{q}_1 \cdot \hat{q}_2)j_{\ell''}(q_1 x_{13}) j_{\ell''}(q_2 x_{23}) Q_{\ell'}(x_{13}, x_{23}) \Big],    
\end{align}
we recovered the real-space expression by also noting that the real-space bispectrum is $B(p_1, p_2, p_3) = B_0(p_1, p_2, p_3) =  B_{00}(p_1, p_2, p_3)/\sqrt{4\pi}$ in our spherical harmonics convention.

%%%%%%%%%%%%%%%%%%%%%%%%%%%%%%%%%%%%%%%%%%%%%%%%%%%%%%%%%%
%%%%%%%%%%%%%%%%%%%%%%%%%%%%%%%%%%%%%%%%%%%%%%%%%%%%%%%%%%
\section{Useful identities}

In this appendix we collect several well-known mathematical identities employed in the derivation of the main results in section~\ref{sec:window_hankel}. Our convention for the spherical harmonics is such that 
\be
Y_{\ell 0}(\theta, \phi) = \sqrt{(2\ell+1)/4\pi} ~ \mathcal{L}_\ell(\theta)\,,
\ee 
where $\mathcal{L}_\ell(\theta)$ is Legendre polynomial of order $\ell$.

\paragraph{Integral representation of Dirac delta distribution:}
\begin{equation}
\label{dirac_delta_integral_rep}
\delta_D(k) = \int \frac{d^3 x}{(2\pi)^3} e^{i \vec{k} \cdot \vec{x}}. 
\end{equation}

\paragraph{Rayleigh expansion of a plane wave:}
\begin{equation}
\label{plane_wave_exp}
    e^{i \vec{k} \cdot \vec{x}} = \sum_\ell i^\ell (2\ell +1) j_\ell(kx) \mathcal{L}_\ell (\hat{k} \cdot \hat{x})\,.
\end{equation}

\paragraph{Addition of spherical harmonics:}
\begin{equation}
\label{addition_spherical_harmonics}
    \mathcal{L}_\ell(\hat{x} \cdot \hat{y}) = \frac{4\pi}{2\ell +1} \sum_m Y_{\ell m} (\hat{x}) Y_{\ell m}^* (\hat{y})\,.
\end{equation}

\paragraph{Orthogonality of Legendre polynomials:} 
\begin{equation}\label{LL_ortho}
    (2\ell +1) \int \frac{d^2 \hat{k}}{4 \pi} \mathcal{L}_\ell(\hat{k} \cdot \hat{x}) \mathcal{L}_{\ell'}(\hat{k} \cdot \hat{y}) = \delta_{\ell \ell'} \mathcal{L}_\ell(\hat{x} \cdot \hat{y})\,
\end{equation}
which implies that 
\begin{equation}\label{integrate_L_ell}
    \int \frac{d^2 \hat{k}}{4\pi} e^{i \vec{k} \cdot \vec{x}} \mathcal{L}_\ell(\hat{k} \cdot \hat{y}) = i^\ell j_\ell(kx) \mathcal{L}_\ell(\hat{x} \cdot \hat{y})\,.
\end{equation}

\paragraph{Orthogonality of spherical harmonics:}
\be
\label{YY_ortho}
    \int d^2 \hat{k}~ Y_{\ell m} (\hat{k}) Y^*_{\ell' m'}(\hat{k}) = \delta_{\ell \ell'} \delta_{m m'}\,.
\ee
which implies that
\be
\label{integrate_Y_ellm}
    \int \frac{d^2 \hat{k}}{4 \pi} e^{i \vec{k} \cdot \vec{x}} Y_{\ell m}(\hat{k}) = i^\ell j_\ell(kx) Y_{\ell m}(\hat{x})\,.
\ee

\paragraph{Gaunt's integral:}
\be
\label{integration_3Y}
    \int d^2 \hat{k}~ Y_{\ell_1 m_1} (\hat{k}) Y_{\ell_2 m_2} (\hat{k}) Y_{\ell_3 m_3} (\hat{k}) = \mathcal{G}^{m_1 m_2 m_3}_{\ell_1 \ell_2 \ell_3}\, ,
\ee
which implies
\begin{equation}\label{integration_3L}
\begin{split}
    \int d^2 \hat{k} ~ \mathcal{L}_{\ell_1}(\hat{k} \cdot \hat{x}) \mathcal{L}_{\ell_2}&(\hat{k} \cdot \hat{y}) \mathcal{L}_{\ell_3}(\hat{k} \cdot \hat{z}) = \\&\sum_{m_1, m_2, m_3} \frac{(4 \pi)^3}{(2\ell_1+1) (2 \ell_2 +1)(2 \ell_3 +1)} \mathcal{G}^{m_1 m_2 m_3}_{\ell_1 \ell_2 \ell_3} Y_{\ell_1 m_1}^*(\hat{x}) Y_{\ell_2 m_2}^*(\hat{y}) Y_{\ell_3 m_3}^*(\hat{z}) \, ,
    \end{split}
\end{equation}
with the Gaunt coefficients are given in terms of Wigner 3$j$ symbols as 
\begin{equation}\label{gaunt}
    \mathcal{G}^{m_1 m_2 m_3}_{\ell_1 \ell_2 \ell_3} = \sqrt{\frac{(2\ell_1+1)(2\ell_2+1)(2\ell_3+1)}{4 \pi}} \begin{pmatrix}
\ell_1 & \ell_2 & \ell_3\\
0 & 0 & 0
\end{pmatrix} \begin{pmatrix}
\ell_1 & \ell_2 & \ell_3\\
m_1 & m_2 & m_3 
\end{pmatrix} \,.
\end{equation}

\paragraph{Product of Legendre polynomials:} 
\begin{equation}\label{contraction_legendre}
    \mathcal{L}_{\ell_1} (\hat{k}\cdot \hat{x}) \mathcal{L}_{\ell_2} (\hat{k}\cdot \hat{x}) = \sum_{\ell_3} \begin{pmatrix}
\ell_1 & \ell_2 & \ell_3\\
0 & 0 & 0
\end{pmatrix}^2 (2\ell_3+1) \mathcal{L}_{\ell_3} (\hat{k}\cdot \hat{x}) \,.
\end{equation}

\paragraph{Product of spherical harmonics:}
\begin{equation}\label{contraction_Y}
    Y_{\ell_1 m_1} (\hat{n}) Y_{\ell_2 m_2} (\hat{n}) = \sum_{\ell_3, m_3} \mathcal{G}_{\ell_1 \ell_2 \ell_3}^{m_1 m_2 m_3} Y^*_{\ell_3 m_3} (\hat{n})\, .
\end{equation}

\paragraph{Orthogonality of Wigner 3$j$ symbols}
\begin{equation}
    \sum_{m_1, m_2} (2\ell_3+1) \begin{pmatrix}
    \ell_1 & \ell_2 & \ell_3 \\
    m_1 & m_2 & m_3
    \end{pmatrix}
    \begin{pmatrix}
    \ell_1 & \ell_2 & \ell'_3 \\
    m_1 & m_2 & m'_3 
    \end{pmatrix}=\delta_{\ell_3 \ell'_3} \delta_{m_3 m'_3}.
\end{equation}

\setlength{\bibsep}{2pt plus 0.5ex}
\bibliographystyle{JHEP}
\bibliography{cosmologia}

\end{document}